\documentclass[fleqn,10pt]{wlscirep}
\usepackage[utf8]{inputenc}
\usepackage[T1]{fontenc}

\usepackage{comment}
\usepackage{hhline}
\usepackage{multirow}
\usepackage{soul} 

\definecolor{2RC}{RGB}{255, 140, 0}
\definecolor{RWC}{RGB}{139, 0, 0}
\definecolor{SFS}{RGB}{220, 20, 60}
\definecolor{ELO}{RGB}{70, 130, 180}
\definecolor{RPR}{RGB}{75, 0, 130}

\newcommand{\tnb}{t_\mathrm{nb}}

\title{Hierarchy and ranking in pairwise sports contests}

\author[1,2]{Bogdán Asztalos}
\author[1]{Boldizsár Balázs}
\author[1,2,*]{Gergely Palla}
\author[1]{Tamás Vicsek}
\affil[1]{Department of Biological Physics, Eötvös Loránd University, Pázmány P. stny. 1/A, H-1117 Budapest, Hungary}
\affil[2]{Health Services Management Training Centre, Semmelweis University, Kutvölgyi út 2, H-1125 Budapest, Hungary}

\affil[*]{gergely.palla@emk.semmelweis.hu}


\begin{abstract}
Ranking athletes by their performance in competitions and tournaments is common in every popular sport and has significant benefits that contribute to both the organization and strategic aspects of competitions. Although rankings are perhaps the most concise and most straightforward representation of the relative strength among the competitors, beyond this one-dimensional characterization, it is also possible to capture the relationships between athletes in greater detail. Following this approach, our study examines the networks between athletes in individual sports such as tennis and fencing, where the nodes are associated with the contestants and the edges are directed from the winner to the loser. We demonstrate that the connections formed through matches arrange themselves into a time-evolving hierarchy, with the top players positioned at its apex. The structure of the resulting networks exhibits detectable differences depending on whether they are constructed purely from round-robin data or from purely elimination-style tournaments. We find that elimination tournaments lead to networks with a smaller level of hierarchy and thus, importantly, to an increased probability of circular win-loss situations (cycles). The position within the hierarchy, along with other network metrics, can be used to predict match outcomes. In the systems studied, these methods provide predictions with an accuracy comparable to that of forecasts based on official sports ranking points or the Elo rating system. A deeper understanding of the delicate aspects of the networks of pairwise contests enhances our ability to model, predict, and optimize the behaviour of many complex systems, whether in sports tournaments, social interactions, or other competitive environments.

\end{abstract}
\begin{document}

\flushbottom
\maketitle

\thispagestyle{empty}

\section*{Introduction}

Quantifying the relative strength of competitors is a fundamental aspect of any sport, traditionally accomplished through linear ranking systems. However, these one-dimensional lists often oversimplify the complex web of relationships and performance dynamics that arise from direct, pairwise competition. Network science offers a robust framework for modelling complex systems by representing entities as nodes and their interactions as edges~\cite{Laci_revmod,Dorog_book,Newman_Barabasi_Watts,Jari_Holme_Phys_Rep,Vespignani_book}. Applying this paradigm to the world of professional sports, where athletes are nodes and match outcomes are directed edges, provides a novel lens to analyse competitive hierarchies and predict future performance.

The present research examines key features of directed networks related to pairwise sports contests within tournament frameworks. 
We will use the terms 'contest' and 'match', as well as 'players' and 'contestants', interchangeably. The topic related to rankings and networks of competitors have recently become popular for interpreting results in the realm of sports tournaments \cite{Bozoki2016, DeBacco2018,zappala2022role,Newman2023,zappala2023paradox}. As for rankings in general, the recent book by P.~Érdi\cite{Erdi2019} has an extensive overview of the many facets and applications of the field. Rankings are usually created by comparing the relevance of the entities considered. A common situation is when the more "important" of the two entities is given a higher score or rank, where importance can correspond to various features. For example, in the case of sports, the winning potential of a contestant is considered a key feature.

In addition to linear rankings, we also analyse the competitive relationships as a hierarchy, which provides a more nuanced, multidimensional ordering of players, with elite performers at the top. In general, hierarchical organisation is a ubiquitous feature of complex networks \cite{Newman_hier,Pumain_book,zafeiris2017why}, observed across a remarkable range of systems, from the intricate regulatory networks within cells \cite{Zeng_Ecoli,hier_methyl} and the social structures of animal groups \cite{Tamas_pigeons,Pigeon_context,Ozogany}, to the organization of online news content \cite{news_portals}, scientific journals and scientific fields \cite{Palgrave,mesh}, and even the grand scale of ecological systems \cite{Hirata_eco,Wickens_eco} and evolution \cite{Eldrege_book,McShea_organism,Mengistu_evolv_hier}.  
In such networks, nodes positioned higher in the hierarchy typically have a greater influence than those at lower levels. Identifying and quantifying this hierarchical structure is a non-trivial challenge, with various approaches ranging from statistical inference based on network topology \cite{Newman_hier} to the development of specific hierarchy measures \cite{Sneppen_hier_measures,mones2012hierarchy,Sole_hier_PNAS,czegel2015random,Gupte_hier_measure,Elisa_hier_measure}. Hierarchies are often represented as directed acyclic graphs that capture asymmetric relationships, such as parent-child or leader-follower, that define the hierarchical ordering of nodes.

In this work, we primarily address the case where two players participating in tournaments are compared based on their prior performance against each other and other contestants. We consider two branches of sports, tennis and fencing, so that the specificity and the generality of our results could be assessed. In addition, for fencing competitions, we analyse data available for the group (round-robin) stage, where each contestant in a group has a bout with every fencer in that group. Associating points or scores with the winning potential of a sports person or a competing individual has a history dating back to the middle of the last century \cite{guttman1946approach, elo1967proposed}. Although the terminology of a contest depends on the branch of sport (e.g., in fencing they are called “bouts”), we shall use the term “matches” for contests of various kinds since this expression applies to the broadest selection of sports.

Due to the ever-increasing interest in tennis, this sport has become a subject of extensive study from both ranking and prediction perspectives. Several authors have also addressed the network aspects of predictions (\cite{Radicchi2011, Bozoki2016, Giorgi2019, Yue2022, arcagni2023new}). To support the applicability of our approach beyond tennis, we also evaluate results published for several variants of fencing bouts. 


Ranking competitors in sports and other competitive fields has several significant benefits 
that contribute to both the organization and strategic aspects of competitions.  Ranking gives a quantifiable measure of a competitor's performance relative to others. This enables competitors, coaches, teams, investors, and sponsors to evaluate their strengths and weaknesses in comparison to their peers.  Rankings are also crucial for creating fair matchups in tournaments. By seeding competitors based on their rankings, organizers can ensure that the strongest competitors do not meet until later in the competition, which maintains competitive interest and fairness throughout the event.  In addition, rankings can serve as a motivational tool for competitors by setting benchmarks and goals. Achieving a higher ranking can be a significant accomplishment, providing recognition and validating the effort and skill of a competitor. From a research perspective, rankings provide valuable data for analysing trends, predicting outcomes, and studying the dynamics of sport at various levels. This can enhance the understanding of how different factors influence competitive success.

A well-known and widely used example for ranking players is the Elo ranking\cite{elo1967proposed} method, which is based on the formulas of the Bradley--Terry model\cite{bradley1952rank} with specific parameters, assuming that the probability of victory of player $i$ over player $j$ is
\begin{equation}
    \mathrm{Pr} (\textrm{$i$ wins over $j$})  = \frac{1}{1 + 10^{(s_j - s_i) / 400}}, \label{eq:bt-model}
\end{equation}
where $s_i$ and $s_j$ are the Elo scores of players $i$ and $j$. As Jerdee and Newman suggested in their work\cite{jerdee2024luck}, this formula can be extended with two parameters:
\begin{equation}
    \mathrm{Pr} (\textrm{$i$ wins over $j$}) = \frac{1}{2} \alpha + (1 - \alpha) \frac{1}{1 + \mathrm{e}^{\beta (s_j - s_i)}} \label{eq:prediction}
\end{equation}
where $\alpha$ quantifies probability of an extreme upset win, and $\beta$ represents the depth of the competition.

By employing advanced analytical tools and methodologies such as centrality measures, motif analysis, and temporal network analysis, one can uncover the hidden features of complex networks such as the networks of matches. Understanding these delicate aspects of networks not only enriches our comprehension of the network's structure and function but also enhances our ability to model, predict, and optimize the behavior of complex systems, whether in sports tournaments, social interactions, or other competitive environments.


\section*{Results}

\subsection*{Data}

We collected a series of match results from various sports disciplines spanning specific periods. During our study, we worked with two sports - tennis and fencing - in which matches consist of a duel of two opposing individuals and end with the clear victory of one of them. Men and women compete separately in these two sports, and in fencing, there are also three disciplines (foil, épée, and sabre), each of which has its own completely independent competition. 
Hence, we have studied 8 different competition categories, which are summarized in Table \ref{tab:datasets}.

Table \ref{tab:datasets} also lists the type of events taken into account. In each category, we have limited the analysis to the most elite events for two reasons: (i) The number of events increases drastically on lower levels of competitions, but due to their decreasing relevance, they are not recorded as rigorously in the freely available databases as the more prestigious ones. (ii) On lower levels of competition the differences between player abilities reduce thus, the reliability of ranking scores loses its significance.

\begin{table}
    \centering
    \begin{tabular}{*2{c}*3{c|}c}
        Sport & Considered events & Discipline & Sex & Number of matches & Source \\\hline
        \multirow{4}{*}{Tennis} & \multirow{4}{*}{\shortstack{Grand Slams \\ Tour-level events \\ Davis Cup \\ \\ from years 1968--2024}} & & \multirow{2}{*}{Men} & \multirow{2}{*}{194,996} & \multirow{2}{*}{Jeff Sackmann\cite{atp-tennis-data}} \\
        & & & & & \\
        & & & \multirow{2}{*}{Women} & \multirow{2}{*}{158,092} & \multirow{2}{*}{Jeff Sackmann\cite{wta-tennis-data}} \\
        & & & & & \\\hline
        \multirow{6}{*}{Fencing} & \multirow{6}{*}{\shortstack{Olympics \\ World Championships \\ Grand Prix \\ World Cups \\ Zone Championships \\ \\ from years 2015--2024}} & \multirow{2}{*}{Foil} & Men & 54,629 & \multirow{6}{*}{Anya Post-Michaelsen\cite{fie-fencing-data}} \\
        & & & Women & 43,798 & \\\hhline{~~---~}
        & & \multirow{2}{*}{Épée} & Men & 73,068 & \\
        & & & Women & 55,290 & \\\hhline{~~---~}
        & & \multirow{2}{*}{Sabre} & Men & 49,458 & \\
        & & & Women & 43,571 & \\\hline
    \end{tabular}
    \caption{Sport result datasets we used for our analysis.}
    \label{tab:datasets}
\end{table}

\subsection*{Time evolving network representation of sport matches}

Based on past match results, we constructed networks where nodes represent players and directed links represent matches, pointing from the winner to the loser. If two players played against each other $n$-times with the same result, we assigned $n$-times more weight to the corresponding link. If the matches of the two players ended with different results, we represented them with two opposite links between the two nodes, each with the appropriate weight. The goal of these networks is to store the complex statistical information gained from past matches to use it to make inferences about the relationship between players. Although all matches in the past carry information, it is not worth keeping all matches from the sport's entire history because their relevance decreases over time: player abilities change, and an overwhelming victory over a much weaker player from several years ago does not indicate that the winner is still that much superior today. Hence, we define a network by taking matches into account only from the previous $\tnb$-long \emph{network-building period}, in which match results presumably reveal something about the players' current abilities. Sliding this network-building time window over time allows us to view the network as a time-dependent network that changes constantly, with its links at time $t$ corresponding to matches from the time interval $[t - \tnb, t]$.  In the following, we use the value of $\tnb = 1\,\mathrm{year}$, in line with the fact that in both tennis and fencing, the official ranking is calculated based on performance over the last year.

\begin{figure}
    \hspace*{-1cm}\includegraphics[width=7.5in]{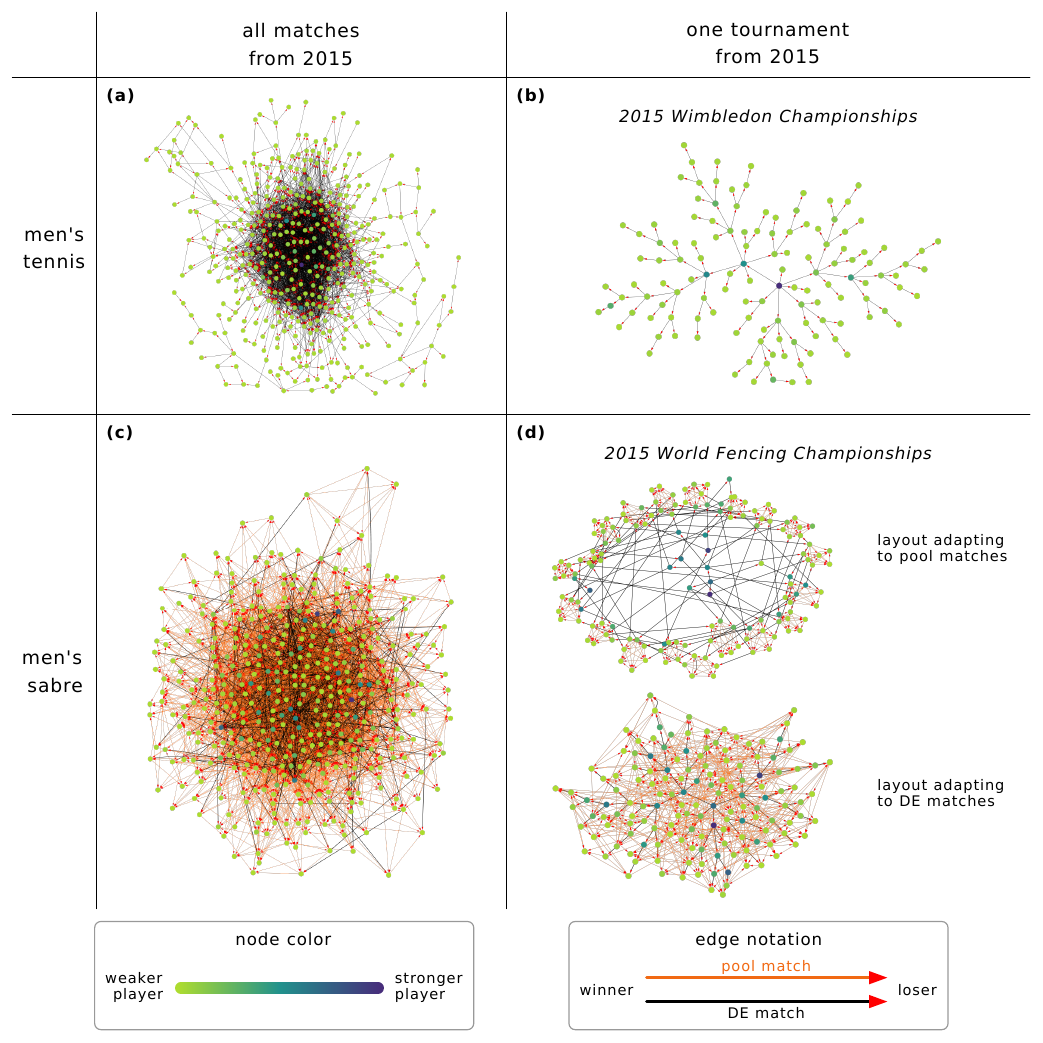}
    \caption{\textbf{Illustration of sport match networks.} In panels \textbf{(a)} and \textbf{(c)}, we plotted the networks constructed from men's tennis and men's sabre matches in the year 2015, as described in the text. Although differences in their structural feature can be observed (the system of relations between nodes of the tennis network in panel (a) is apparent, while the relatively large number of pool matches makes the fencing network in panel (c) more complex), the core of both networks is too tangled and crowded for the human eye. Hence, in panels \textbf{(b)} and \textbf{(d)}, we plotted the networks gained based on only one tournament. The organizational difference between the two sports is apparent: the tennis tournament is in a round-robin format, while the fencing tournament has two phases: a pool (group) phase and a direct elimination (DE) phase. In panel (d), there are two instances of the same network, differing only in their layout; one highlights the structure of the pool phase, and the other highlights the structure of the DE phase of the tournament.}
    \label{fig:networks}
\end{figure}

For an illustration, we plotted a network of tennis matches and another network of fencing matches from one single year in Figure \ref{fig:networks}. Besides, we also plotted in Figure \ref{fig:networks} two networks, one based on a tennis tournament and one based on a fencing tournament, visualising the difference between their organisation method.

\subsection*{Elements of hierarchy in the network of matches}

It is natural to assume a hierarchy among players in any individual sport, similar to a dominance hierarchy, where the best players are positioned at the top of the hierarchy, and weaker players are placed lower. Our network construction is well-suited to capture this organization, as in the idealized scenario of stronger players consistently winning, the resulting networks become directed acyclic graphs, perfectly reflecting a hierarchical structure. On the contrary, in reality, stronger players cannot always win: surprising results and upset wins are not uncommon in sports, so the resulting graph will not be perfectly acyclic. However, the expectation regarding sports is still that the match results reflect a kind of dominance, and the network defined above should show some hierarchical nature.


\subsubsection*{Quantifying hierarchy in the network of matches}

\begin{figure}
    \centering
    \includegraphics[width=6in]{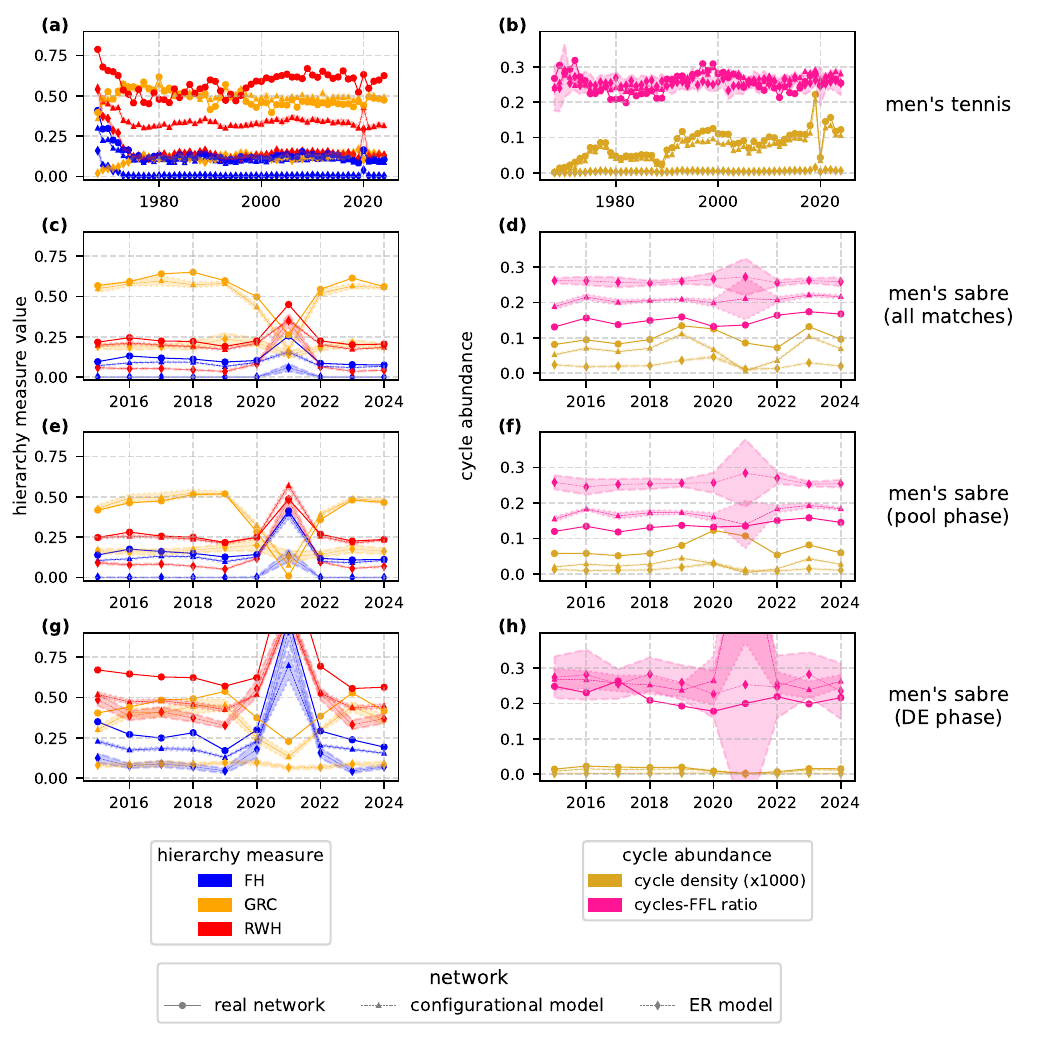}
    \caption{\textbf{Hierarchy measures and 3-cycle abundance 
    in the network of matches.} \textbf{(a)} The flow hierarchy (blue), the global reaching centrality based on 2-reach (orange), and the random walk hierarchy (red) for the network of men's tennis matches over time, displaying both the measured value in the original networks (circles) and also the average value in randomised networks according to the configuration model (triangles) and the Erdős--Rényi model (diamonds). The shaded regions around the averages indicate the standard deviation.
    \textbf{(b)} The number of 3-cycles in the graph normed by the number all node triads (yellow), and its ratio to the number of feedforward 3-loops (FFLs) (pink) for the network of men's tennis matches over time, displaying both the measured value in the original networks (circles), and also the average value in randomised networks according to the configuration model (triangles) and the Erdős--Rényi model (diamonds). The shaded regions around the averages indicate the standard deviation.
    \textbf{(c)}-\textbf{(d)} Same as (a)-(b) but for the network of men's sabre matches. \textbf{(e)}-\textbf{(f)} Same as (a)-(b) but for the network of matches played in the pool phase in men's sabre tournaments. \textbf{(g)}-\textbf{(h)} Same as (a)-(b) but for the network of matches played in the direct elimination phase in men's sabre tournaments.}
    \label{fig:compare-hierarchy}
\end{figure}

To examine the extent to which the constructed networks exhibit signs of hierarchy, we employed three different hierarchy measures that capture distinct intuitions of hierarchy. In the case of flow hierarchy (FH) \cite{luo2011detecting} the fraction of links not taking part in any cycles is calculated, the global reaching centrality (GRC) \cite{mones2012hierarchy} is based on the inhomogeneity of the reaching centrality distribution, and the random walk hierarchy measure (RWH) \cite{czegel2015random} uses a random walk process for discovering potential leaders. (A detailed description of the used hierarchy measures is given in the Methods.)
In Figure \ref{fig:compare-hierarchy}a and c, we plot these hierarchy measures over time for men's tennis and men's sabre. Here, we also compare the observed values with the results obtained in networks randomized according to the configuration model and the Erdős-Rényi model. (The analogous plots for the other categories are provided in the Supplementary Information.)

Apparently, in the case of men's tennis (Figure \ref{fig:compare-hierarchy}a), there is a large gap between the hierarchy measure of the original network and that of an Erdős--Rényi network (having the same number of nodes and links) for all years and all three hierarchy measures. In the meantime, comparisons with the configuration network ensemble yield mixed outcomes. The original networks exhibit higher FH and RWH values across all years, with a substantial difference in RWH. Conversely, the original network's GRC values were either comparable to or, in some years, lower than the ensemble average. In the case of men's sabre (Figure \ref{fig:compare-hierarchy}c), we can observe a large gap between the hierarchy measures of the original networks and the average hierarchy in the corresponding Erdős--Rényi random networks, similarly to what we found for the men's tennis networks. For the configuration model-based randomisation, the average hierarchy is again lower than the original value across all years and all hierarchy measures; however, this time, the difference compared to the original value is smaller in magnitude. 

For both of these networks, we can observe a large impact of the Covid pandemic in the years 2020-2021. Due to the lockdown, the organisation of the tournaments, as well as the participation of the players in tournaments, was severely affected, resulting in anomalous peaks in our analysis.

The comparison between the results for men's tennis networks and the men's sabre networks is inconclusive: the RWH values are higher for tennis, the GRC values are higher for sabre, and the FH values are roughly equal. This incomparability can potentially be explained by the differences in the implementation of tournaments between the two sports. Conventionally, tennis tournaments are organised in a single-elimination format, where the loser of each match is eliminated from the tournament, thereby creating a tree-shaped subgraph of the participants. This means that the winner of each tournament is the dominant node of the tournament's sub-graph, and the overall structure of the whole network will consist of these sub-trees, dominated by the few best players who have won (or almost won) all the tournaments they started in. On the contrary, fencing tournaments begin with a \emph{pool phase}, where participants are divided into pools of 5-7 players, and they play against all the other players in the pool, aiming to earn the right to proceed to the \emph{direct elimination phase}. This format facilitates the formation of cycles in the graph, as players in the same pool will inevitably create a weakly linked clique, thereby undermining the network's hierarchical structure.
Consequently, to get a clearer picture of hierarchy within these sports, we need to separate the following two effects and watch them on their own: 1) the difference between the hierarchy of the round-robin format and the direct elimination format, and 2) the hierarchy that appears itself independently on the organisational structure (i.e., the degree of dominance between players).

To examine the difference between the round-robin format and the direct elimination format, we directly compared the pool phase and the direct elimination phase of sabre tournaments directly to each other. In Figure \ref{fig:compare-hierarchy}e and g, we recreate the plots of panels Figure \ref{fig:compare-hierarchy}c for networks built of matches only from the pool phase and the direct elimination phase, respectively. FH and RWC values are firmly larger for the direct elimination network, but so are they for the networks generated with the configuration model, suggesting that this is due to the tree-shaped skeleton of the graph. Let's compare the hierarchy measures of real networks to those of homogeneous Erdős-Rényi networks. The ratio is larger for all three measures in the pool phase networks, which can be interpreted as indicating that the superior-subordinate relationships among pool matches were more decisive.

To test the conjecture that the level of hierachy of the networks of games depends on the way the tournaments are organized, we examined directly the number of cycles in the networks. Within an ideal hierarchical structure of edges, the directions of links determine a one-way route from dominant nodes to subordinate ones, and the possibility of returning from a subordinate node to a dominant one would imply the breaking of the hierarchy. Hence, the lack of circular node triads i.e., paper-rock-scissors-like relationships between players, is a clear trace of player dominance. In Figure \ref{fig:compare-hierarchy}b, d, f, and h, we measure how abundantly 3-cycles are present in the studied graphs. The cycle abundance is calculated by comparing the number of 3-cycles in the graphs to all possible node-triads in the graph (cycle density), or to all weakly linked node-triads, including 3-cycles and feedforward 3-loops (cycle-FFL ratio). (The detailed description of the cycle abundance measures is given in the Methods.) Apparently, for networks of tennis matches (Figure \ref{fig:compare-hierarchy}b) and for networks of the DE phase of sabre tournaments (Figure \ref{fig:compare-hierarchy}h) the real cycle-FFL ratio does not differ significantly from the randomised networks (neither in the case of the configuration model, nor in the case of Erdős--Rényi model), but for networks where pool matches contribute, the cycle abundance is lower. This suggests that player strength dominance does not stand out in the single-elimination format, and the degree of hierarchy we observe in this case stems from the actual structure of the network.

\begin{figure}
    \centering
    \includegraphics[width=4.6in]{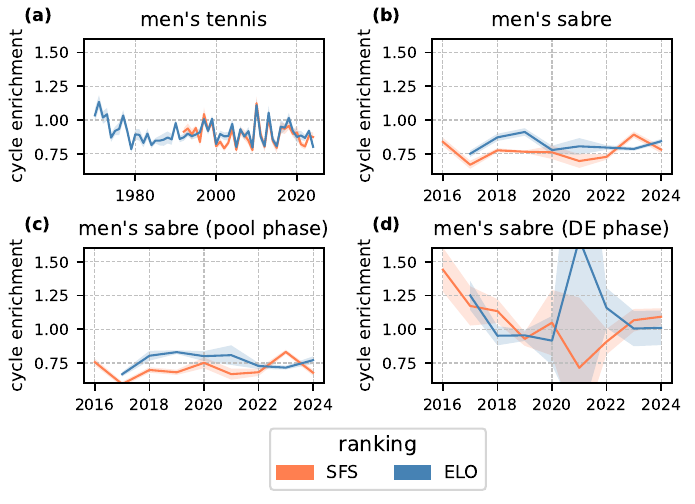}
    \caption{\textbf{Cycle enrichment calculated with SFS and ELO rankings in the network of matches.} The fraction of the number of 3-cycles and the expected number of 3-cycles from a given player strength distribution. The shaded regions around the averages indicate the standard deviation. \textbf{(a)} The cycle enrichment values for the network of men's tennis matches over the years. \textbf{(b)} The cycle enrichment values for the network of men's sabre matches over the years. \textbf{(c)} The cycle enrichment values for the network of matches played in the pool phase in men's sabre tournaments. \textbf{(d)} The cycle enrichment values for the network of matches played in the direct elimination phase in men's sabre tournaments.}
    \label{fig:cycle-enrichment}
\end{figure}

We also compared the number of 3-cycles in networks to their expected number with a given strength distribution. In the following subsections of the Results, we explain how player strengths can be measured using traditional methods or the concept of hierarchy, and how outcome probabilities can be estimated accordingly. If we resample the result of each match based solely on the probabilities coming from the strength difference of players, the number of cycles can be considered as an expected value following the specific dominance and organization situation. If we observe fewer cycles in the networks of played matches, the hierarchy of players has a more substantial effect on the real system. However, if there is an enhanced number of cycles, it means that more upset wins occurred, and the assumed dominance between players could not prevail. We measured the enrichment of the network with 3-cycles by calculating the fraction of cycles and the expected value using a given ranking score. (The detailed description of the cycle enrichment measure is given in the Methods.) In Figure \ref{fig:cycle-enrichment}, we plot this enrichment value over time with two different ranking scores for both men's tennis and men's sabre. (The analogous plots for the other categories and ranking scores are provided in the Supplementary Information.)
They give roughly the same results: for both tennis and sabre data, the enrichment value is smaller than 1 in Figure \ref{fig:cycle-enrichment}a and b, with sabre values being slightly smaller, suggesting that the hierarchy between sabre fencers has larger relevance. In Figure \ref{fig:cycle-enrichment}c and d, we can see that the cycle enrichment values are significantly larger for DE matches, which means that in the elimination format, dominance is less effectively present, and more upset wins happen between players.



\subsubsection*{Structure of hierarchy}

One of the most essential features of a hierarchical system is that a few superior agents dominate over many subordinate ones. This means the system must have a widening shape from top to bottom, as there are more units on the lower levels of the hierarchy than on the upper levels. However, the exact structure of this widening can depend on several factors. The simplest case of hierarchy is when each superior dominates over the same number of subordinates, which results in an exponential multiplication of the number of nodes from level to level, but real hierarchical organisations can show different structures depending on the specific relational structure behind it. In the following, we examine if the networks of sports matches have such a widening shape.

\begin{figure}[p]
    \centering
    \includegraphics[width=\textwidth]{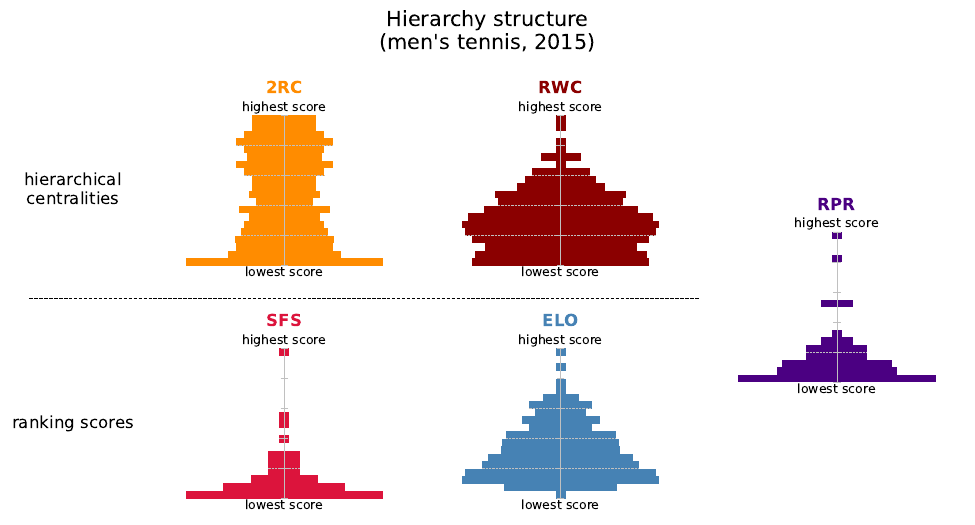}
    \caption{\textbf{The structure of the hierarchy for men's tennis.} The distribution of nodes along the 2-reach centrality (orange), the random walk centrality (red), the official sport ranking (pink), the Elo score (cyan), and the revered pagerank (purple) for the network of men's tennis matches played in 2015. The width of the bars represents the relative number of nodes having a specific centrality value on log-scale.}
    \label{fig:atp-tennis-hists}
\end{figure}

\begin{figure}[p]
    \centering
    \includegraphics[width=\textwidth]{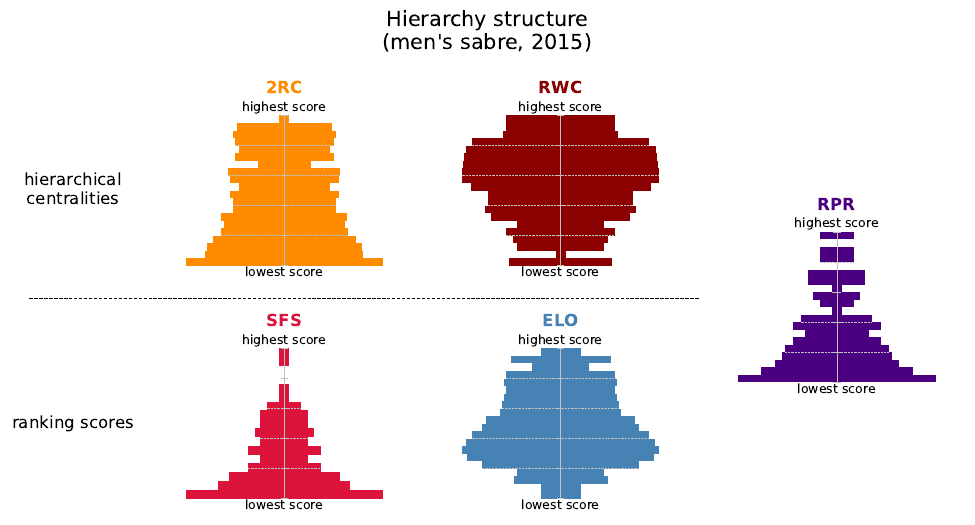}
    \caption{\textbf{The structure of the hierarchy for men's sabre.} The distribution of nodes along the 2-reach centrality (orange), the random walk centrality (red), the official sport ranking (pink), the Elo score (cyan), and the revered pagerank (purple) for the network of men's sabre matches played in 2015. The width of the bars represents the relative number of nodes having a specific centrality value on log-scale.}
    \label{fig:men-sabre-hists}
\end{figure}

To measure the widening of the hierarchy across levels, we first need to define a method to measure the level of each separate node and determine the distribution of nodes according to this quantity. Due to the intuitive character of hierarchy, defining the position of nodes within it can also not be done universally, and different measures can be found along different approaches. We used 2 different node centralities (or local measures) to capture how dominant their role is within the hierarchical structure of the network, the 2-reach centrality (2RC) \cite{borgatti2003key}, counting what portion of all nodes of the network can be reached within two steps from the given node, and the random walk centrality (RWC) \cite{czegel2015random}, taking the probability that a random walker discovering potential leaders stays on the given node. (A detailed description of the used hierarchy measures is given in the Methods). In the upper row of Figures \ref{fig:atp-tennis-hists} and \ref{fig:men-sabre-hists}, the distribution along these two quantities is shown for men's tennis and men's sabre in a vertical view visualising the widening of the hierarchical structure (analogous plots for other sport disciplines are provided in the Supplementary Information). Although a certain extent of widening can be observed in some cases, there is not a universal pattern for the structures of the network structure along the hierarchical centralities.

\subsection*{Ranking}

\subsubsection*{Ranking scores and hierarchical centralities}

The most important and most interesting question in any sport is who the absolute best player is, or in a more general way, what is the order of players based on their abilities. Similarly to hierarchy, playing ability is also an intuitive notion, and finding a formal definition or method of measurement for it can be done in several different ways depending on what kinds of performance we consider as the signs of the absolute greatness. Nevertheless, due to the cultural needs of fans, most sports have an official ranking system, scoring player's performance according to certain criteria which reflect the sport's legacy in a more or less objective way, so in this paper, we use the official sports federation's ranking score (SFS) for each sport as one of the studied ranking scores. Besides, we also calculated the  Elo score (ELO) for each player based on the matches in the database, since it is widely viewed as one of the most reliable ranking score. It is essential to note that the databases used do not contain all matches of the sport. The Elo scores we counted are not the actual Elo scores of the players; they can be used only for comparison within this elite group of players and matches. 

Let us notice that besides these two classical direct performance-awarding score systems, any quantity can be interpreted as ranking score that generally assigns higher values to players who win more matches. Hierarchical centralities, for example, measure the role of nodes (players) within the hierarchical structure of the network of matches, which can be developed by gaining victories, hence hierarchical centralities should also be able to serve as ranking scores. This can be justified by the intuition of sport dominance having similar attributes to the dominance in a social relation system. Conversely, classical ranking scores are also suitable to examine the structure of hierarchy since the distribution of players along a given score can counted independently on the network itself as shown in Figures \ref{fig:atp-tennis-hists} and \ref{fig:men-sabre-hists}.

As an illustration of the similar but not identical intuition behind different ranking scores, Tables \ref{tab:tennis-ranking} and \ref{tab:sabre-ranking} present the top 5 players according to the various ranking scores for men's tennis and men's sabre in 2015. It can be observed that the specific order varies through the separate lists, some particularity appears almost everywhere, for example, Novak Djokovic was the undisputable best tennis player in that year according to four out of the five rankings, or {\'A}ron Szil{\'a}gyi and Junghwan Kim were among the best 5 sabre player according to three out of the five rankings.

\begin{table}
    \centering

    \begin{tabular}{c|cc}
        \multicolumn{3}{c}{\textcolor{2RC}{\textbf{2RC}}} \\
        rank & player & score [$\cdot 10^4$] \\\hline
        1 & Novak Djokovic & 5561 \\
        2 & Andy Murray    & 5444 \\ 
        3 & Rafael Nadal   & 5421 \\
        4 & Stan Wawrinka  & 5374 \\
        5 & John Isner     & 5304   
    \end{tabular}
    \hspace{2cm}
    \begin{tabular}{c|cc}
        \multicolumn{3}{c}{\textcolor{RWC}{\textbf{RWC}}} \\
        rank & player & score [$\cdot 10^6$] \\\hline
        1 &Marcelo Arevalo  & 9469 \\
        2 &Marton Fucsovics & 8985 \\ 
        3 &Mohammad Ghareeb & 7636 \\
        4 &Chieh Fu Wang    & 7545 \\
        5 &Patrick Tierro   & 6917   
    \end{tabular}\\
    \vspace{0.5cm}
    \begin{tabular}{c|cc}
        \multicolumn{3}{c}{\textcolor{SFS}{\textbf{SFS}}} \\
        rank & player & score \\\hline
        1 & Novak Djokovic & 16585 \\
        2 & Andy Murray    & 8945  \\ 
        3 & Roger Federer  & 8265  \\
        4 & Stan Wawrinka  & 6865  \\
        5 & Rafael Nadal   & 5230      
    \end{tabular}
    \hspace{2cm}
    \begin{tabular}{c|cc}
        \multicolumn{3}{c}{\textcolor{ELO}{\textbf{ELO}}} \\
        rank & player & score \\\hline
        1 & Novak Djokovic & 2416 \\
        2 & Roger Federer  & 2267 \\ 
        3 & Andy Murray    & 2189 \\
        4 & Rafael Nadal   & 2134 \\
        5 & Stan Wawrinka  & 2071        
    \end{tabular}\\
    \vspace{0.5cm}
    \begin{tabular}{c|cc}
        \multicolumn{3}{c}{\textcolor{RPR}{\textbf{RPR}}} \\
        rank & player & score [$\cdot 10^5$] \\\hline
        1 & Novak Djokovic & 6125 \\
        2 & Roger Federer  & 5197 \\ 
        3 & Andy Murray    & 3360 \\
        4 & Stan Wawrinka  & 3114 \\
        5 & Rafael Nadal   & 2133        
    \end{tabular}
    \caption{The first 5 players of men's tennis according to different local measures as ranking score in 2015.}
    \label{tab:tennis-ranking}
\end{table}

\begin{table}
    \centering

    \begin{tabular}{c|cc}
        \multicolumn{3}{c}{\textcolor{2RC}{\textbf{2RC}}} \\
        rank & player & score [$\cdot 10^4$] \\\hline
        1 & Luca Curatoli & 7635 \\
        2 & Bongil Gu     & 7190 \\ 
        3 & Matyas Szabo  & 7166 \\
        4 & Renzo Agresta & 7119 \\
        5 & Bolade Apithy & 7119   
    \end{tabular}
    \hspace{2cm}
    \begin{tabular}{c|cc}
        \multicolumn{3}{c}{\textcolor{RWC}{\textbf{RWC}}} \\
        rank & player & score [$\cdot 10^6$] \\\hline
        1 & Dmytro Raskosov      & 3375 \\
        2 & Francesco d'Armiento & 3361 \\ 
        3 & Kamil Ibragimov      & 3351 \\
        4 & Luigi Samele         & 3334 \\
        5 & Aron Szilagyi        & 3320   
    \end{tabular}\\
    \vspace{0.5cm}
    \begin{tabular}{c|cc}
        \multicolumn{3}{c}{\textcolor{SFS}{\textbf{SFS}}} \\
        rank & player & score \\\hline
        1 & Junghwan Kim     & 243 \\
        2 & Aron Szilagyi    & 219  \\ 
        3 & Vincent Anstett  & 181  \\
        4 & Alexey Yakimenko & 169  \\
        5 & Nikolay Kovalev  & 151      
    \end{tabular}
    \hspace{2cm}
    \begin{tabular}{c|cc}
        \multicolumn{3}{c}{\textcolor{ELO}{\textbf{ELO}}} \\
        rank & player & score \\\hline
        1 & Nicolas Limbach    & 1813 \\
        2 & Junghwan Kim       & 1790 \\ 
        3 & Bongil Gu          & 1782 \\
        4 & Tiberiu Dolniceanu & 1771 \\
        5 & Max Hartung        & 1769        
    \end{tabular}\\
    \vspace{0.5cm}
    \begin{tabular}{c|cc}
        \multicolumn{3}{c}{\textcolor{RPR}{\textbf{RPR}}} \\
        rank & player & score [$\cdot 10^5$] \\\hline
        1 & Kamil Ibragimov & 1969 \\
        2 & Bongil Gu       & 1883 \\ 
        3 & Max Hartung     & 1751 \\
        4 & Aron Szilagyi   & 1749 \\
        5 & Junghwan Kim    & 1662        
    \end{tabular}
    \caption{The first 5 players of men's sabre according to different local measures as ranking score in 2015.}
    \label{tab:sabre-ranking}
\end{table}

Although the above hierarchical centralities can be considered as ranking scores, their definition was adapted to a network-based point of view. Hierarchical centralities are typically designed for systems, where knowing the domination status of different agents has a global advantage (e.g. animal groups should know who the leader is, companies should have a clear subordination organisation). In sports, however, the apparent dominance of a single person does not have a social profit, in fact, the opposite: if one player always wins, the sport becomes boring. Hence, it would be worth using a network quantity that measures social dominance directly, being relevant in sports, instead of mathematical intuitions of network hierarchy.

For this, let us consider the \emph{prestige} gained by winning a match as something illustrious and triumphant, that is worth chasing. If one catches this \emph{prestige}, their goal is not to lose it. Suppose that if a player loses a match, the \emph{prestige} in the possession of this player goes to the winner. If a defeated player wants to regain some \emph{prestige}, they must win matches again. Also, \emph{prestige} is fleeting, so it decreases in time and restarts its journey among players from time to time, so players with high reputations need to continue winning to keep their status. In this picture, \emph{prestige} can be described as a walker on the network of players which steps in the direction of victories. Arcagni et al. \cite{arcagni2023new} interpreted this as the match being a skill transfer from the loser to the winner, and the player's ability should be measured by an eigenvalue centrality in a directed network, where the direction of links must be reversed compared to our definition. In this paper, we used the simplest eigenvalue centrality, the PageRank algorithm, designed by the founders of Google, Brin and Page \cite{brin1998anatomy}. Figures \ref{fig:atp-tennis-hists} and \ref{fig:men-sabre-hists} also show the distribution of nodes along reversed PageRank (RPR). In the following, we use this as a fifth local measure, which is both a hierarchical network centrality and a ranking score.

\subsubsection*{Decay of ranking score}

\begin{figure}[p]
    \centering
    \includegraphics[width=\textwidth]{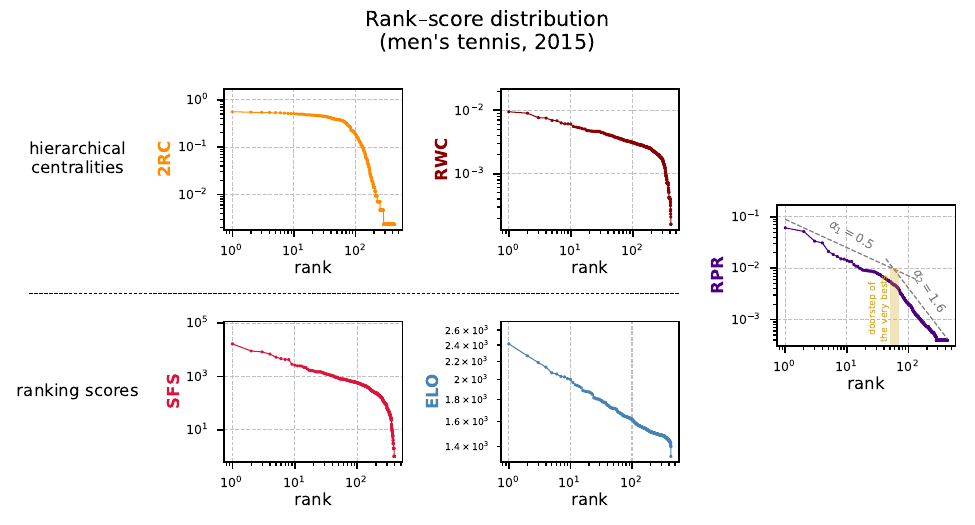}
    \caption{\textbf{The rank--score distribution of local measures for men's tennis.} The rank--score distribution of nodes along the 2-reach centrality (orange), the random walk centrality (red), the official sport ranking (pink), the Elo score (cyan), and the reverse pagerank (purple) for the network of men's tennis matches played in 2015. In the case of reversed pagerank two segments can be found following power-law scaling.}
    \label{fig:atp-tennis-zipf}
\end{figure}

\begin{figure}[p]
    \centering
    \includegraphics[width=\textwidth]{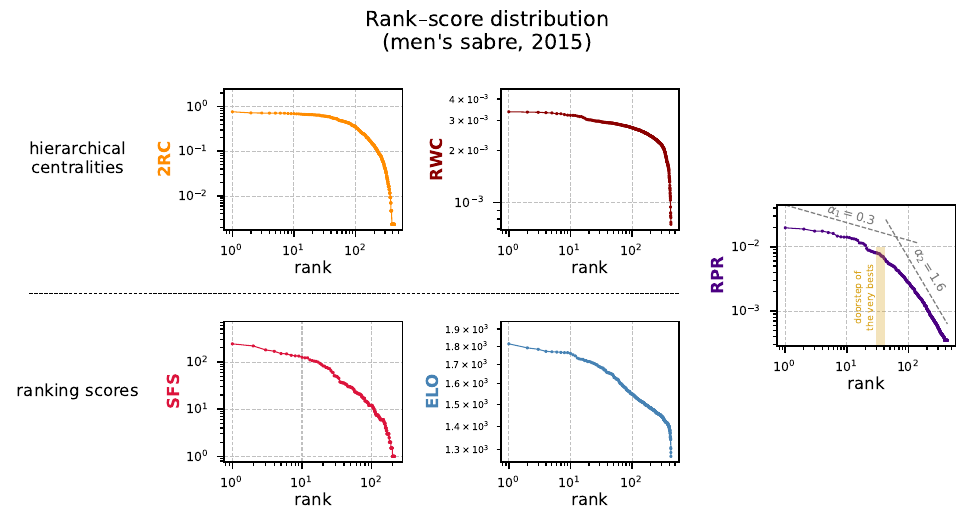}
    \caption{\textbf{The rank-score distribution of local measures for men's sabre.} The rank--score distribution of nodes along the 2-reach centrality (orange), the random walk centrality (red), the official sport ranking (pink), the Elo score (cyan), and the revered pagerank (purple) for the network of men's sabre matches played in 2015. In the case of reversed pagerank two segments can be found following power-law scaling.}
    \label{fig:men-sabre-zipf}
\end{figure}

Since each player has several local measures, it can be asked how they relate to each other. Is the best player an outstanding talent, or is there a close competition for the first place? Are there more people with lower score values, or does the density of players remain the same at each level of competition? To answer these questions, we plotted the rank--score distribution of the different ranking scores in Figure \ref{fig:atp-tennis-zipf} and \ref{fig:men-sabre-zipf} for men's tennis and men's sabre, respectively (analogous plots for other sport disciplines are provided in the Supplementary Information). It can be generally observed that on a log-log scale, hierarchical centralities (2RC and RWC) start with a long plateau and fall sharply, ELO values all are in the same order of magnitude, and the decay of SFS shows some concave curve. The distribution of RPR values, on the other hand, shows power-law decay through more than one order of magnitude for all studied sports disciplines, and in most cases, more segments can be observed realising power-law decay with different power exponents. Similar phenomena occur in other fields of complex systems where power-law distributions are the consequence of the inner interaction of the system, and the breaking point of the curve marks a transition point characteristic to the effect 
Our findings suggest that RPR is suitable for measuring hierarchy from the perspective of such a complex interactive effect and for identifying the point that separates the very best players from other elite athletes according to this effect.

\subsection*{Prediction using score differences}

As mentioned above, a ranking score records the order of dominance between players and suggests that a player with a higher rank in a pairing should win the match. However, in sports, upset wins often occur, sometimes due to pure luck but sometimes because better rank does not indicate being in better shape under the given circumstances. Thus, a ranking score does not tell us everything about a future sports event just provides us with a hint about who is more likely to win. Hence, the quality of a ranking system can be characterised by how accurately estimates the probability of the different outcomes of a match. A good example of this is the Bradley--Terry model\cite{bradley1952rank} which is explicitly designed in such a way that the exponential ranking scores must be proportional to the chance of victory, and the difference between opponents gives a tangible estimate for the probability of the outcomes through formula (\ref{eq:bt-model}). Since the rank order of players can be composed based on any quantity defined for players, a natural question is how well certain quantities predict match outcomes. In this section, we examine all the 5 local measures defined above in the light of prediction and compare their accuracy to each other.

We used formula (\ref{eq:prediction}) for calculating winning probabilities, and fitted the optimal parameters $\hat{\alpha}_s (t)$ and $\hat{\beta}_S (t)$ in each year $t$ for each score then, we used again formula (\ref{eq:prediction}) with the parameter values $\alpha = \hat{\alpha}_s (t)$ and $\beta = \hat{\beta}_S (t)$ to predict result of next year ($t + 1$) matches. The details of the fitting and prediction procedure are explained in the Methods section. We calculated the outcome probabilities of all matches which had enough prior information in all 8 competition categories with all 5 ranking scores and compared the probability estimates to the real results of the matches, which we know so in retrospect. We measured the accuracy of different prediction methods through the fraction of mistakes (FM), the average squared error (SE), and the average linear error (LE) of the calculated probabilities (see the formulas in Methods). In Table \ref{tab:prediction}, the accuracies of different prediction scores are shown among the other competition categories using the different ranking scores. For tennis, ELO performs the best, but for fencing disciplines, SFS overachieves according to the most accurate measures. This means that the official ranking method in fencing is more efficient than in tennis, whose official ranking method underachieves even some of the hierarchy measures.

\begin{table}
    \centering
    \begin{tabular}{*3{c|}c}
        \multicolumn{4}{c}{\textbf{Men's tennis}} \\
        centrality & FM & SE & LE \\\hline
        2RC & 0.345 & 0.219 & 0.437 \\
        RWC & 0.450 & 0.245 & 0.489 \\
        SFS & 0.356 & 0.224 & 0.447 \\
        ELO & \textbf{0.344} & \textbf{0.213} & \textbf{0.426} \\
        RPR & 0.345 & 0.217 & 0.433
    \end{tabular}
    \hspace{2cm}
    \begin{tabular}{*3{c|}c}
        \multicolumn{4}{c}{\textbf{Women's tennis}} \\
        centrality & FM & SE & LE \\\hline
        2RC & 0.347 & 0.219 & 0.436 \\
        RWC & 0.437 & 0.243 & 0.485 \\
        SFS & 0.350 & 0.223 & 0.442 \\
        ELO & \textbf{0.339} & \textbf{0.209} & \textbf{0.417} \\
        RPR & 0.344 & 0.215 & 0.429
    \end{tabular}
    \vspace{0.6cm}
    
    \begin{tabular}{*3{c|}c}
        \multicolumn{4}{c}{\textbf{Men's sabre}} \\
        centrality & FM & SE & LE \\\hline
        2RC & 0.320 & 0.209 & 0.408 \\
        RWC & 0.374 & 0.229 & 0.457 \\
        SFS & \textbf{0.272} & \textbf{0.194} & 0.386 \\
        ELO & 0.297 & 0.194 & \textbf{0.384} \\
        RPR & 0.317 & 0.207 & 0.414
    \end{tabular}
    \hspace{2cm}
    \begin{tabular}{*3{c|}c}
        \multicolumn{4}{c}{\textbf{Women's sabre}} \\
        centrality & FM & SE & LE \\\hline
        2RC & 0.347 & 0.221 & 0.435 \\
        RWC & 0.400 & 0.235 & 0.468 \\
        SFS & \textbf{0.296} & \textbf{0.205} & 0.411 \\
        ELO & 0.320 & 0.205 & \textbf{0.407} \\
        RPR & 0.340 & 0.217 & 0.435
    \end{tabular}
    \vspace{0.6cm}
    
    \begin{tabular}{*3{c|}c}
        \multicolumn{4}{c}{\textbf{Men's épée}} \\
        centrality & FM & SE & LE \\\hline
        2RC & 0.360 & 0.224 & 0.445 \\
        RWC & 0.412 & 0.239 & 0.475 \\
        SFS & \textbf{0.314} & \textbf{0.213} & \textbf{0.428} \\
        ELO & 0.343 & 0.215 & 0.432 \\
        RPR & 0.359 & 0.225 & 0.449
    \end{tabular}
    \hspace{2cm}
    \begin{tabular}{*3{c|}c}
        \multicolumn{4}{c}{\textbf{Women's épée}} \\
        centrality & FM & SE & LE \\\hline
        2RC & 0.329 & 0.213 & 0.417 \\
        RWC & 0.371 & 0.231 & 0.455 \\
        SFS & \textbf{0.278} & 0.198 & 0.394 \\
        ELO & 0.304 & \textbf{0.197} & \textbf{0.390} \\
        RPR & 0.326 & 0.210 & 0.421 \\
    \end{tabular}
    \vspace{0.6cm}

    \begin{tabular}{*3{c|}c}
        \multicolumn{4}{c}{\textbf{Men's foil}} \\
        centrality & FM & SE & LE \\\hline
        2RC & 0.327 & 0.210 & 0.416 \\
        RWC & 0.377 & 0.230 & 0.461 \\
        SFS & \textbf{0.278} & \textbf{0.197} & \textbf{0.395} \\
        ELO & 0.306 & 0.197 & 0.396 \\
        RPR & 0.325 & 0.209 & 0.423
    \end{tabular}
    \hspace{2cm}
    \begin{tabular}{*3{c|}c}
        \multicolumn{4}{c}{\textbf{Women's foil}} \\
        centrality & FM & SE & LE \\\hline
        2RC & 0.329 & 0.213 & 0.417 \\
        RWC & 0.371 & 0.231 & 0.455 \\
        SFS & \textbf{0.278} & 0.198 & 0.394 \\
        ELO & 0.304 & \textbf{0.197} & \textbf{0.390} \\
        RPR & 0.326 & 0.210 & 0.421
    \end{tabular}
    
    \caption{Prediction accuracy in different sports disciplines using different centralities as the "prediction score". The best accuracy value is highlighted in bold for each category and accuracy measure. The best score to use for prediction is usually ELO or SFS, but this strongly depends on the sport and the method used to measure error.}
    \label{tab:prediction}
\end{table}

\section*{Discussion}

In the highly competitive world of professional tennis, public perception often gravitates towards a well-defined hierarchy of elite players. Dominant figures in the sport, frequently spotlighted in the media, are perceived as nearly invincible against opponents with lower ATP rankings. This traditional view suggests a predictable landscape in which top-ranked players consistently outperform their lesser-ranked counterparts. However, our analysis of the network of matches tells a strikingly different story.

Utilizing a network-based approach, where nodes represent players and directed edges indicate match outcomes from winner to loser, we uncover an unexpected pattern in the data. Despite the clear hierarchy suggested by ATP rankings and public perception, the frequency of lower-ranked players defeating their higher-ranked peers is not only non-trivial but surprisingly high. This observation is exemplified by the significant number of cycles within our network, constructed from matches among the top 1000 ranked players in a given year.
This paradox challenges the conventional wisdom of player dominance in tennis and illuminates the intricate dynamics of competition that standard rankings fail to capture. Such findings not only enhance our understanding of sports competitions but also provide valuable insights into the nature of hierarchy and performance in any competitive organization. By revealing the underlying complexity and unpredictability of outcomes, our study encourages a reevaluation of how success and potential are assessed in competitive fields.

The implications extend beyond sports, suggesting that in any hierarchical organization, apparent stability and predictability might mask a dynamic interplay of factors that could lead to unexpected outcomes. Thus, our work contributes to a deeper understanding of the elements that govern success and the potential for disruption in established hierarchies.

When analysing networks of matches we draw attention to a crucial aspect of competitive sports and other similar systems where pairwise elimination is used to determine a winner. The points we raised are as follows.
{\it Hierarchical Nature of Tournaments}: Elimination tournaments, where each round halves the number of participants, inherently creates a pyramid-like hierarchy where the fewest compete the most. This method of elimination ensures that the final rounds are contested by those who have survived multiple rounds of competition, which might suggest they are the strongest or most skilled participants. {\it Impact on Scoring and Ranking}: The point about accumulating scores or points based on the number of matches played is particularly relevant. Players who advance further in the tournament naturally have more opportunities to score points, which could skew rankings toward those who play more matches, rather than those who might be more skilled but were eliminated early due to the tournament draw or early tough matchups. {\it Enhanced Hierarchy from Tournament Design}: We propose that the hierarchical organization of tournaments results in an enhanced hierarchy of points is a plausible interpretation. It suggests that tournaments don't just find the best player but also amplify the perceived difference in skill or strength between tiers of players based on the cumulative scoring system. This can be seen as both a feature and a bug of tournament design—effective for determining a winner but potentially misleading in evaluating a player's true strength across the board. {\it Broader Implications}: This analysis is relevant not only to sports but also to any competitive field that employs a similar elimination-based structure to rank participants, such as gaming, business competitions, or academic contests. It raises important questions about the fairness and accuracy of such systems and whether they truly identify the 'best' or simply the 'last standing'.

Our arguments encourage a deeper examination of how we design competitive systems and what modifications might be needed to ensure they are fair and truly reflective of individual abilities and strengths. This could lead to exploring alternative tournament formats that might offer a more balanced approach to competition and ranking.

\section*{Methods}

\subsection*{Hierarchical centralities and hierarchy measures}

The intuition of a hierarchical network is that there are a few dominant nodes which have superior relationships with the others from some point of view. Since this intuition can be formalised in many different ways, different approaches have been used to describe the hierarchy in networks. Some of them are applicable both on a global level, characterising the hierarchy of the entire network, and on a local level, characterising the role of individual nodes within a hierarchical structure. In this paper, we discuss both levels.

On local level, we used the following hierarchical centralities:
\begin{description}
    \item[m-reach centrality (mRC)] is the number of nodes which can be reached in $m$ steps starting from a certain node. It can be viewed as extensions of node degree since degree is the number of direct neighbours of the certain node, so it corresponds to $m = 1$ \cite{borgatti2003key}. In this paper, we used $m = 2$ (noted by 2RC), and normed the value with the number of all reachable nodes of the network:
    $$
        \mathrm{2RC} (i) = \frac{1}{N - 1} \cdot \# \{\textrm{reachable nodes from $i$ in $2$ steps}\}
    $$
    where $N$ is the size of weakly connected component in which node $i$ is present.
    
    \item[random walk centrality (RWC)] is based on a random walk model on the network which is represent the information flow over the links. The walking model is defined by the transition matrix $T_{ij}$ describing the probability of the walker going to node $i$ from $j$, which depends on the degree of nodes $i$ and $j$, and an arbitrary parameter $f$, which describes the characteristic distance made by the walker. According to the considerations of Czégel and Palla\cite{czegel2015random}, the stationary probability distribution $\boldsymbol{p}^{\mathrm{stat}}$ of the walker is given by the eigenvalue problem
    $$
        \boldsymbol{p}^{\mathrm{stat}} = \frac{1}{1 + f} \left[ \boldsymbol{T} \left( \boldsymbol{p}^{\mathrm{stat}} + \frac{f}{N} \boldsymbol{1} \right) \right]
    $$
    where $N$ is the number of nodes. The specific probability values can be viewed as a centrality measure of nodes:
    $$
        \mathrm{RWC} (i) = (\boldsymbol{p}^{\mathrm{stat}})_i.
    $$
\end{description}

On global level, we used the following hierarchy measures:
\begin{description}
    \item[Flow hierarchy (FH)] is defined as the fraction of links which do not participate in any cycles, i.e.,
    $$
        \mathrm{FH} = \frac{1}{L} \cdot \sum_{i \in E} e_i
    $$
    where $E$ denotes the set of links, $L$ is the number of links (i.e., $L = |E|$), and $e_i = 0$ if link $i$ is in a cycle while $e_i = 1$ if it is not. This quantity measures the extent to which a network's structure resembles a directed acyclic graph, where information or control flows unidirectionally without cycles \cite{luo2011detecting}.
    
    \item[Global reaching centrality (GRC)] can be defined as the heterogeneous distribution of an m-reach centrality. Using $m = 2$, if $\mathrm{2RC} (i)$ is the 2-reach centrality of node $i$ and $\mathrm{2RC}^{\max}$ is the highest centrality value, then following the formalism of Mones et al.\cite{mones2012hierarchy}, GRC is defined as
    $$
        \mathrm{GRC} = \frac{1}{N - 1} \cdot \sum_{i \in V} \left[ \mathrm{2RC}^{\max} - \mathrm{2RC} (i) \right]
    $$
    where $V$ is the set of nodes and $N$ is the number of nodes (i.e., $N = |V|$).

    \item[Random walk hierarchy (RWH)] is based on a similar idea as GRC, but using the random walk centrality. Since the distribution $\mathrm{RWC}$ values characterises how important the specific nodes in terms of information flow, its relative width
    $$
        \mathrm{RWH} = \sigma (\boldsymbol{p}^{\mathrm{stat}})\,/\,\mu (\boldsymbol{p}^{\mathrm{stat}})
    $$
    measures the degree of hierarchy of the network itself \cite{czegel2015random}.
\end{description}

\subsection*{Ranking methods}

Player ability is also an abstract concept, and quantifying it can be done in many different ways. The situation is also complicated by the fact that an official ranking method of a sport is often not applicable to any other sport, and hence is incomparable. In this paper, we used three universally usable ranking scores to measure player strengths, one of which was the official one, so the comparability was handled through its officialness.

\begin{description}
    \item[Sports federation score (SFS)] is the score calculated by the governing body of the sport, and which is used for determining the "official ranking" of players. Both tennis and fencing have their point system to reward player performance, 
    and a real-time updating list where the current ranking can be followed \cite{atprankings, wtarankings, fierankings}. Due to the officialness of this score, players take this one the most seriously and prioritise to maximise it.

    \item[Elo rating system (ELO)] is calculated by the method suggested by Arpad Elo originally for measuring a player's performance in chess\cite{elo1967proposed}. Since Elo scores are not tracked officially in most sports, we needed to calculate it for the data according to Elo's method: all players started with a score of 1500, and we went through the matches chronologically; when a player with score $s_1$ won against an opponent with score $s_2$, their score changed by $\Delta s_1 = \frac{K}{1 + 10^{(s_1 - s_2) / S}}$ and $\Delta s_2 = - \frac{K}{1 + 10^{(s_1 - s_2) / S}}$, where $K$ and $S$ were chosen to be $S = 400$ and $K = 32$.

    \item[Reversed PageRank (RPR)] is calculated from the networks of matches, based on the concept introduced in the Results: We reversed the direction of links (i.e., as if they pointed from the loser towards the winner) and calculated the PageRank centrality of the nodes. PageRank ranks nodes (originally web pages on the internet) based on the importance of the nodes linking to them\cite{brin1998anatomy}. Nodes with more incoming links from important nodes receive higher PageRank scores, indicating their greater significance in the network. Since in our case the links were reversed, nodes with links towards important nodes received high scores i.e., players who defeated other players with high scores. The exact scores are calculated by solving the following eigenvalue problem:
    $$
        \mathrm{RPR} (i) = \frac{1 - d}{N} + d \cdot \sum_{j \in M (i)} \frac{\mathrm{RPR} (j)}{L (j)}
    $$
    where $N$ is the number of all nodes in the network, $M (i)$ is the set of nodes to which $i$ is linked, $L (j)$ is the number of inbound links on $j$, and $d$ is and arbitrary parameter which we chosen to be $d = 0.85$.
\end{description}

One can see that all of the above scores change over time, thus making it possible for them to reflect the changes in the relationship of player strengths. SFS is rolling both in the tennis and fencing categories with a one-year time window. This means the score of a player is counted from the results achieved in the last 12 months, and when a match becomes older than one year, the points gained by participating in it cease (and possibly replaced by the event of the actual year). The dynamic behaviour of ELO is encoded into its calculating method: each match updates the current score of the players, so the newest match always has the largest effect on the scores. Since RPR is calculated from a time-dependent network, its value changes in the same way as the network itself, similar to the cases of local hierarchy centralities.

\subsection*{Network randomisation methods}

To compare the values of hierarchy measures in real networks, we randomised the original networks and used them as the basis for comparison. We used two randomisation models, which are intended to remove a significant amount of the statistical correlations within the network while preserving its most basic features, allowing the new network to be comparable to the original one. Thus, since the new networks are not completely random but generated from the original one to some extent, their hierarchical structure might also be dependent on the original structure. This must be taken into account during the comparison, and the interpretation of the results must be assessed with due caution, as we explained in the Results. The two randomisation methods are the following:

\begin{description}
    \item[Erdős--Rényi model] keeps the number of nodes and links of the original network while rewriting the links, i.e.\ both source and target nodes are chosen randomly for each link. This means all nodes are equivalent, their degree distribution follows a binomial distribution, and most of the node features have a typical value around which the individual values fluctuate. Thus, an Erdős--Rényi graph is typically not hierarchical and can serve as a standard of the comparison for the first try. \cite{erdos1959random}
    \item[Configuration model] also keeps the number of nodes and links but also reserves the degree distribution of nodes. This is achieved instead of randomly assigning source and target nodes to links, relinking them by switching the targets of two randomly chosen links at once and repeating this switching step many times. (In our case, for a network with $L$ links, we applied $5L$ iterations, so each link was relinked 5 times on average.) Since the degree distribution is guaranteed to be the same as the one of the (supposedly) hierarchical network, there will certainly be some structural differences between nodes, which can easily appear through the value of the hierarchical measures. But this is not the result of the relationship of the nodes but their different state in the network. \cite{bollobas1980probabilistic}
\end{description}

\subsection*{Making predictions using score differences}

In the Results we calculated the probability of outcomes for each match using formula (\ref{eq:prediction}) which contains two parameters $\alpha$ and $\beta$. For optimal prediction, we need to determine their optimal value. Since the temporal character of the data is essential in our problem, we decided to handle the optimal parameter values as temporal variables as well. This is also in accordance with the real-world situation, where bookmakers and gambles try to make predictions knowing the latest results and events.

\begin{figure}
    \centering
    \includegraphics[width=80mm]{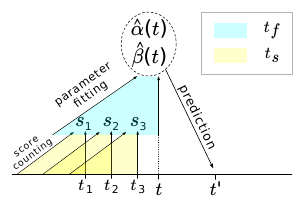}
    \caption{\textbf{Illustration of the fitting process.} We use matches from the score-counting period $t_\mathrm{s}$ (yellow) before certain times ($t_1$, $t_2$, $t_3$) to calculate prediction scores ($s_1$, $s_2$, $s_3$) then, we use such scores from the fitting period $t_\mathrm{f}$ (blue) before time $t$ to find the optimal parameter values $\hat{\alpha} (t)$ and $\hat{\beta} (t)$ which gives us the best trial prediction at $t$. Finally, in the possession of the fitted parameters, we can predict future matches at $t' > t$.}
    \label{fig:fit}
\end{figure}

For fitting on formula (\ref{eq:prediction}) and finding the optimal values $\hat{\alpha} (t)$ and $\hat{\beta} (t)$ at time $t$, we need a certain number of finished matches from before $t$ based on which we calculate the scores, and the trial probabilities for matches at $t$. If we already know the results of matches at $t$, we can determine the best parameter values which result the best trial probabilities and which then can be used for predictions for times $t' > t$. Hence, for making a real prediction for a specific sports event, we need to know the results at least from a previous fitting period $t_{\mathrm{f}}$ on which we fit the parameters on the matches and from a score-counting period $t_{\mathrm{s}}$ before the fitting period which we use for making trial predictions. This relationship between different time windows is illustrated in Figure \ref{fig:fit}.

For scores that are node centralities, the score-counting period is the period from which the matches are used to build the network, i.e. $t_{\mathrm{s}} = t_{\mathrm{nb}}$. For the results shown in Table \ref{tab:prediction}, the length of both time windows were chosen to be $t_{\mathrm{f}} = 1\,\mathrm{year}$ and $t_{\mathrm{s}} = 1\,\mathrm{year}$. Our time resolution was also 1 year, i.e. the scores at the end of the previous year were taken as the actual scores of players through a whole calendar year. (E.g., Novak Djokovic had 11,360 ATP points on December 31, 2014, so we used this value for making predictions for all the matches Djokovic played in 2015.) 

Once, we calculated the outcome probabilities for the matches with a certain score, we compared them with the real results. If we estimated probability $p_m$ for the real outcome of match $m$, the larger $p_m$ value is, the more accurate our method is. We measured the collective accuracy for all $M$ matches of the method by three quantities:

\begin{description}
    \item[Fraction of mistakes (FM)] is the number of matches where we predicted lower probability for the winner to win divided by the number of all matches. Basically, if we have had to guess for each match which player would win, this is the fraction of times we would have been wrong. Formally,
    $$
        \textrm{FM} = \frac{1}{M} \cdot \# \left\{ m : p_m < 0.5 \right\}.
    $$
    (If the estimated probability was 0.5 for both players, we counted it as a half mistake, since in this case we would have to choose between the two players randomly.) Optimization for this quantity can provide us to be the best prediction reliability. For an absolute random prediction, this quantity would be 0.5.
    \item[Average squared error (SE)] is the squared difference between the estimated probability and 1, averaged over all matches, i.e.
    $$
        \textrm{SE} = \frac{1}{M} \cdot \sum_m (1 - p_m)^2.
    $$
    This is the cost function of the fitting using the method of least squares, so optimization for this quantity can be considered as the maximum likelihood strategy. For an absolute random prediction, this would be 0.25.
    \item[Average linear error (LE)] is the linear difference between the estimated probability and 1, averaged over all matches, i.e.
    $$
        \textrm{LE} = \frac{1}{M} \cdot \sum_m (1 - p_m)
    $$
    Since the single difference of $(1 - p_m)$ is the expected value of making a wrong prediction for match $m$, the whole sum can be considered as the risk of betting on the more probable player at each match. Bookmakers and gamblers want to minimize this value, so optimization for this value yields us the best financial profit. For an absolute random prediction, this would be 0.5.
\end{description}

\subsection*{Cycle abundance and cycle enrichment}

To quantify network hierarchy resulting from the organizational structure of the championships and the dominance relationship between players, we presented several method to measure how much more cycle is contained by the graph than it should be in some kinds of idealisation. In Figure \ref{fig:compare-hierarchy}b, d, f, and h, we plotted the cycle abundance measured in two different ways, which are meant to quantify how many cycles the graphs contain normed to their size. The two formalisation of cycle abundance are:
\begin{description}
    \item[Cycle density] is the number of 3-cycles divided by the number of all possible node-triads in the graph, i.e.\ $\binom{N}{3}$ where $N$ is the number of nodes. This is a simple size-normalised number of cycles.
    $$
        \textrm{Cycle density} = \frac{\# \{\textrm{3-cycles}\}}{\binom{N}{3}}
    $$
    \item[Cycle-FFL ratio] is the number of 3-cycles divided by the number of all weakly connected node-triads (i.e.\ the sum of 3-cycles and feedforward 3-loops) in the graph. The normalisation of this quantity is done not by just the size of the graph, but the clustering of nodes.
    $$
        \textrm{Cycle-FFL ratio} = \frac{\# \{\textrm{3-cycles}\}}{\# \{\textrm{3-cycles}\} + \# \{\textrm{FFLs}\}}
    $$
\end{description}

In Figure \ref{fig:cycle-enrichment}, we plotted the cycle enrichment, which measures how many more cycles there are in the network than expected from the given distribution of player abilities. For this, we relinked the network in the following way: we went through all the matches represented by the network's links and guessed the winner of each match by randomly drawing from the Bernoulli distribution given by the winning probability of the prediction process detailed above. When the randomly drawn winner matched the actual winner, we left the link unchanged. However, if the randomly drawn winner was the other player, we reversed the direction of the link. Hence, in the new network, the number of 3-cycles could differ from the original one, while the weak connectivity of nodes remained the same. The ratio of the real number of 3-cycles to their number in the relinked network expresses how much more probable a cycle to form in real life than expected, and how much more enriched the real graph with cycles than the sampled one.
$$
    \textrm{Cycle enrichment} = \frac{\# \{\textrm{3-cycles in the real network}\}}{\# \{\textrm{3-cycles in the simulated network}\}}
$$

\section*{Data availability}

We used datasets publicly available on the internet. Men's and women's tennis data was collected by Jeff Sackmann who made them downloadable from his corresponding GitHub repositories\cite{atp-tennis-data,wta-tennis-data}. All fencing data was downloaded from the official website of the International Fencing Federation, using the code of Anya Post-Michaelsen\cite{fie-fencing-data}.

\section*{Acknowledgements}

This project was supported by the National Research, Development and Innovation Office under grant no. SNN139598, and by the EKÖP-24 University Excellence Scholarship Program of the Ministry fro Culture and Innovation from the Source of the National Research, Development and Innovation Fund.
The authors also thank Péter Érdi and Péter Pollner for the several useful discussions during the work on the project. 

\section*{Author contributions statement}

The concept of the study and the initial results come from B.B., while the detailed analysis of data was performed by B.A. under the supervision of P.G. and T.V. All authors participated in writing and reviewing the manuscript.

\label{lastpage}


\newpage 
\setcounter{figure}{0}
\renewcommand{\thefigure}{S\arabic{figure}}

\vskip-36pt%
{
    \raggedright\sffamily\bfseries\fontsize{20}{25}\selectfont
    Hierarchy and ranking in pairwise sports contests:\\
    Supplementary Information\par
}
\vskip10pt
{
    \raggedright\sffamily\bfseries\fontsize{12}{16}\selectfont 
    Bogdán Asztalos, Boldizsár Balázs, Gergely Palla, and Tamás Vicsek\par
}
\vskip18pt%

In this document, we present the results for each sports category, which could not be included in the main text. In Figures 2-7 of the main text, only the cases of men's tennis and men's sabre are showed, and conclusion is drawn based on them. Here, the analogous figures are collected for all studied sports, and the reader can see that they also support the claims of the paper.
\vspace{2cm}

\section*{Hierarchy measures in the network of matches}

\begin{figure}[h]
    \centering
    \includegraphics[width=6in]{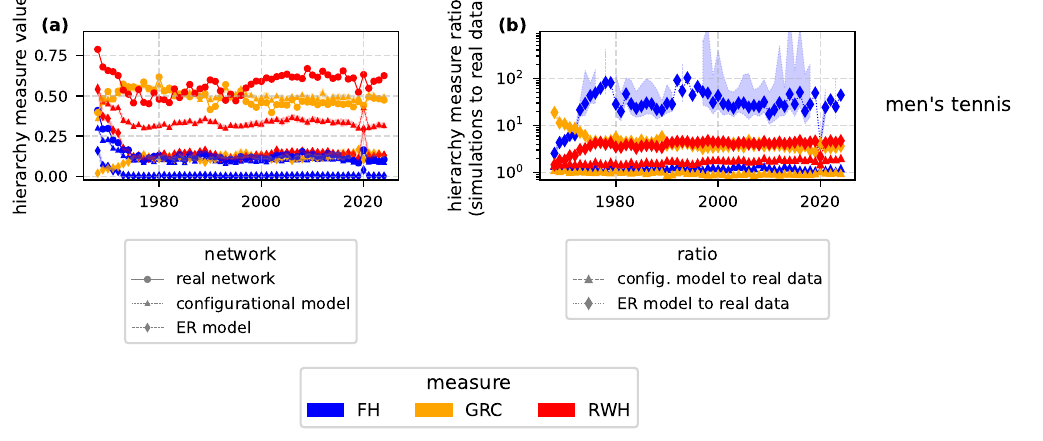}
    \caption{\textbf{Hierarchy measures in the network of men's tennis matches.} \textbf{(a)} The flow hierarchy (blue), the global reaching centrality based on 2-reach (orange), and the random walk hierarchy (red) over time, displaying both the measured value in the original networks (circles) and also the average value in randomised networks according to the configuration model (triangles) and the Erdős--Rényi model (diamonds). \textbf{(b)} The three hierarchy values measured in the real network shown in panel (a), divided by the values in the randomised networks according to the configuration model (triangles) and the Erdős--Rényi model (diamonds).}
    \label{fig:hierarchy-1}
\end{figure}

In the left column of  Figure 2 of the main text, we plotted the different hierarchy measures over time for the network of real matches, and the value of these measures for two types of network randomisation. There, we presented the results of men's tennis tournaments, and men's sabre tournaments. In this section, we show the analogous results of all the studied dataset in Figures \ref{fig:hierarchy-1}-\ref{fig:hierarchy-2}.

\begin{figure}[h]
    \centering
    \includegraphics[width=6in]{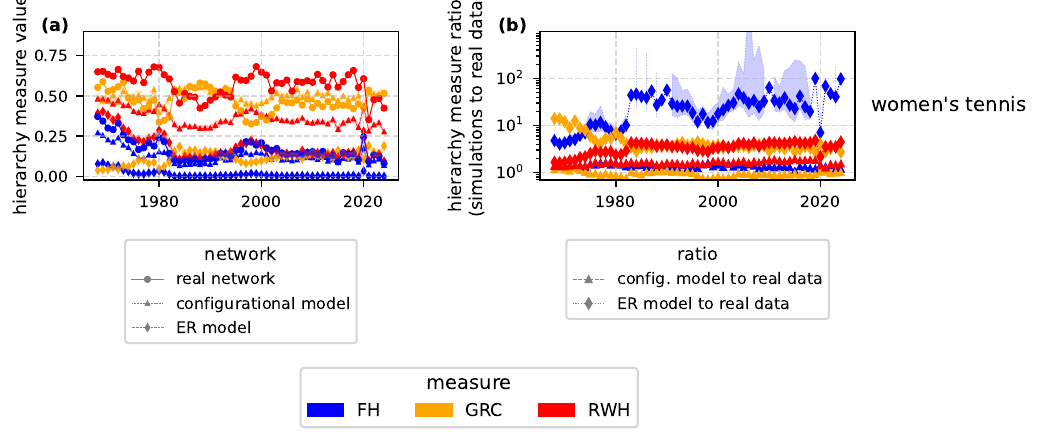}
    \caption{\textbf{Hierarchy measures in the network of women's tennis matches.} \textbf{(a)} The flow hierarchy (blue), the global reaching centrality based on 2-reach (orange), and the random walk hierarchy (red) over time, displaying both the measured value in the original networks (circles) and also the average value in randomised networks according to the configuration model (triangles) and the Erdős--Rényi model (diamonds). \textbf{(b)} The three hierarchy values measured in the real network shown in panel (a), divided by the values in the randomised networks according to the configuration model (triangles) and the Erdős--Rényi model (diamonds).}
\end{figure}

\begin{figure}[h]
    \centering
    \includegraphics[width=6in]{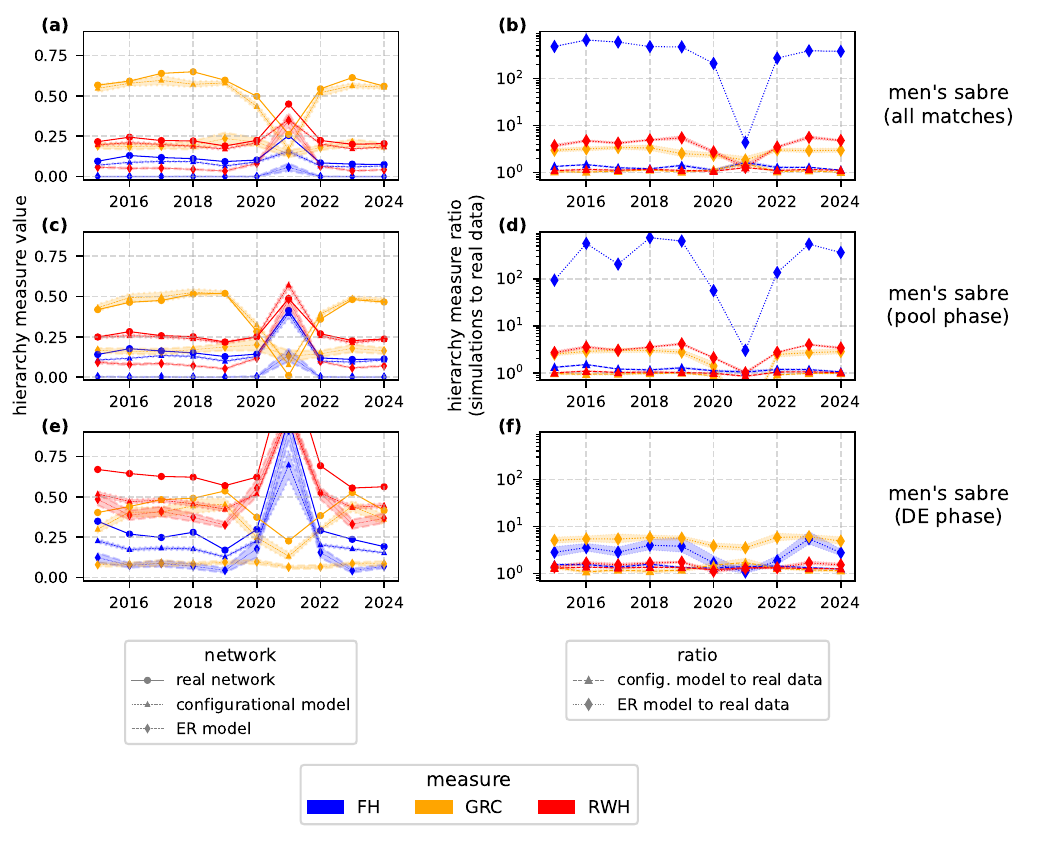}
    \caption{\textbf{Hierarchy measures in the network of men's sabre matches.} \textbf{(a)} The flow hierarchy (blue), the global reaching centrality based on 2-reach (orange), and the random walk hierarchy (red) over time, displaying both the measured value in the original networks (circles) and also the average value in randomised networks according to the configuration model (triangles) and the Erdős--Rényi model (diamonds). \textbf{(b)} The three hierarchy values measured in the real network shown in panel (a), divided by the values in the randomised networks according to the configuration model (triangles) and the Erdős--Rényi model (diamonds).}
\end{figure}

\begin{figure}[h]
    \centering
    \includegraphics[width=6in]{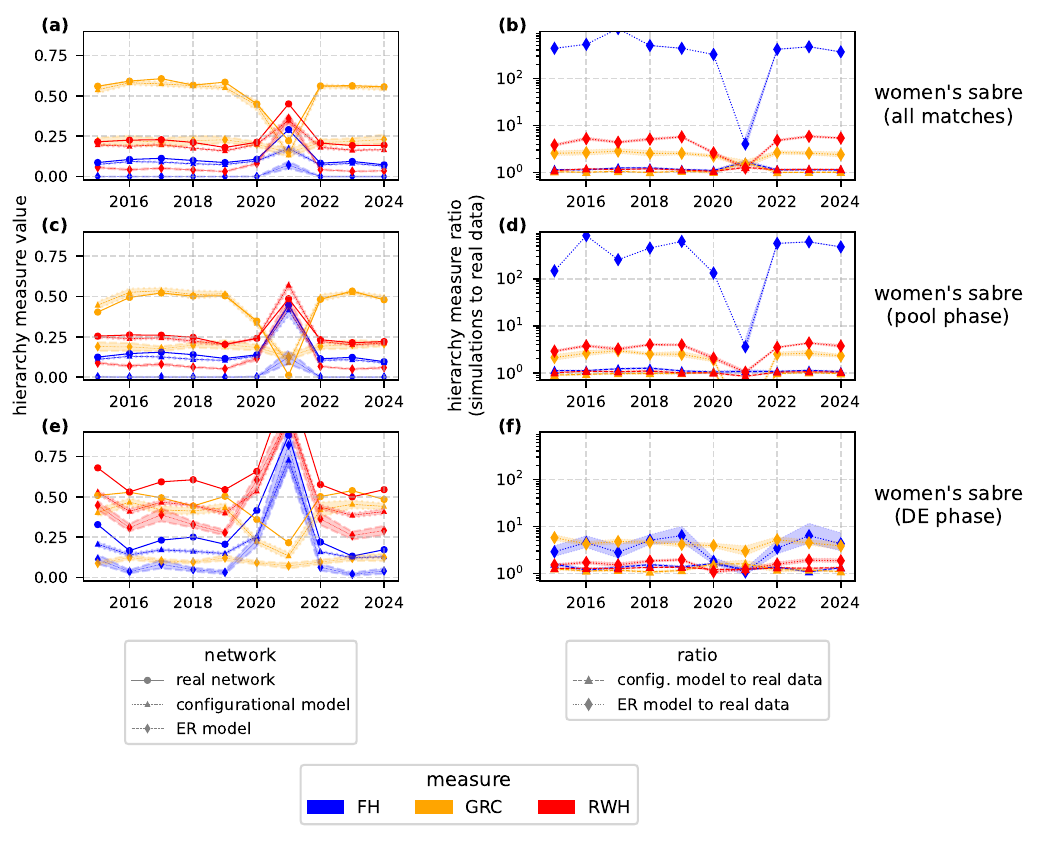}
    \caption{\textbf{Hierarchy measures in the network of women's sabre matches.} \textbf{(a)} The flow hierarchy (blue), the global reaching centrality based on 2-reach (orange), and the random walk hierarchy (red) over time, displaying both the measured value in the original networks (circles) and also the average value in randomised networks according to the configuration model (triangles) and the Erdős--Rényi model (diamonds). \textbf{(b)} The three hierarchy values measured in the real network shown in panel (a), divided by the values in the randomised networks according to the configuration model (triangles) and the Erdős--Rényi model (diamonds).}
\end{figure}

\begin{figure}[h]
    \centering
    \includegraphics[width=6in]{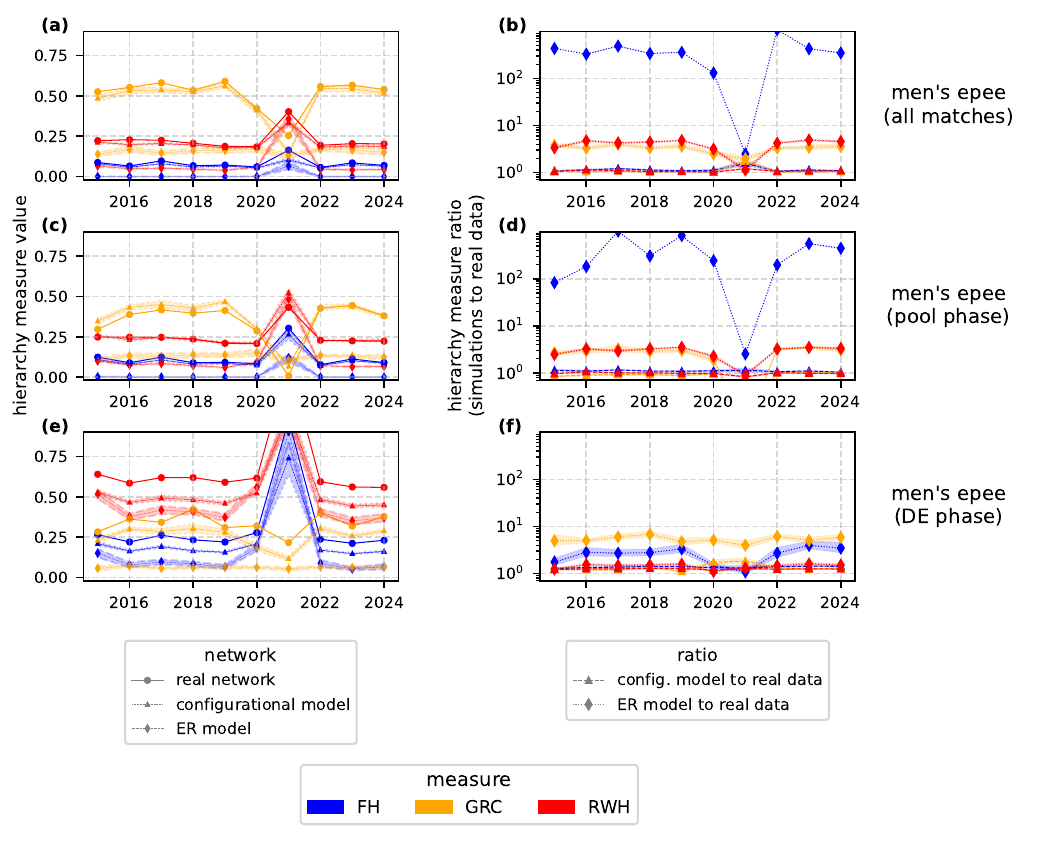}
    \caption{\textbf{Hierarchy measures in the network of men's épée matches.} \textbf{(a)} The flow hierarchy (blue), the global reaching centrality based on 2-reach (orange), and the random walk hierarchy (red) over time, displaying both the measured value in the original networks (circles) and also the average value in randomised networks according to the configuration model (triangles) and the Erdős--Rényi model (diamonds). \textbf{(b)} The three hierarchy values measured in the real network shown in panel (a), divided by the values in the randomised networks according to the configuration model (triangles) and the Erdős--Rényi model (diamonds).}
\end{figure}

\begin{figure}[h]
    \centering
    \includegraphics[width=6in]{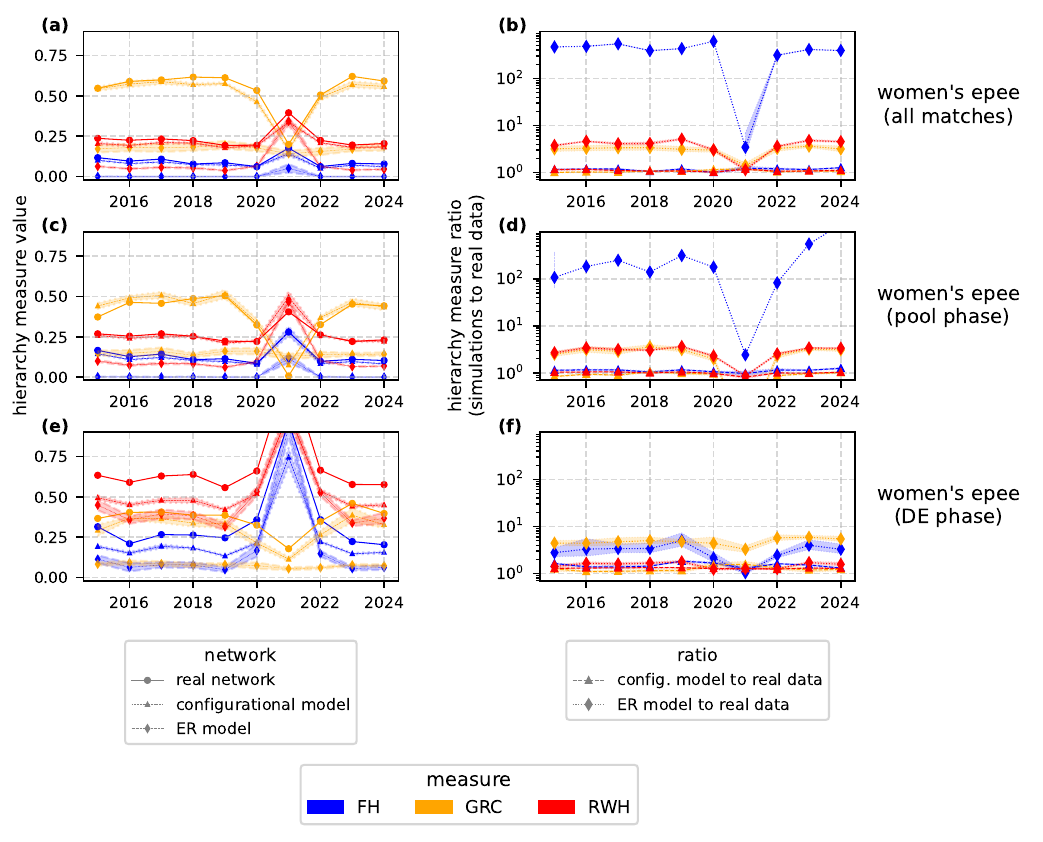}
    \caption{\textbf{Hierarchy measures in the network of women's épée matches.} \textbf{(a)} The flow hierarchy (blue), the global reaching centrality based on 2-reach (orange), and the random walk hierarchy (red) over time, displaying both the measured value in the original networks (circles) and also the average value in randomised networks according to the configuration model (triangles) and the Erdős--Rényi model (diamonds). \textbf{(b)} The three hierarchy values measured in the real network shown in panel (a), divided by the values in the randomised networks according to the configuration model (triangles) and the Erdős--Rényi model (diamonds).}
\end{figure}

\begin{figure}[h]
    \centering
    \includegraphics[width=6in]{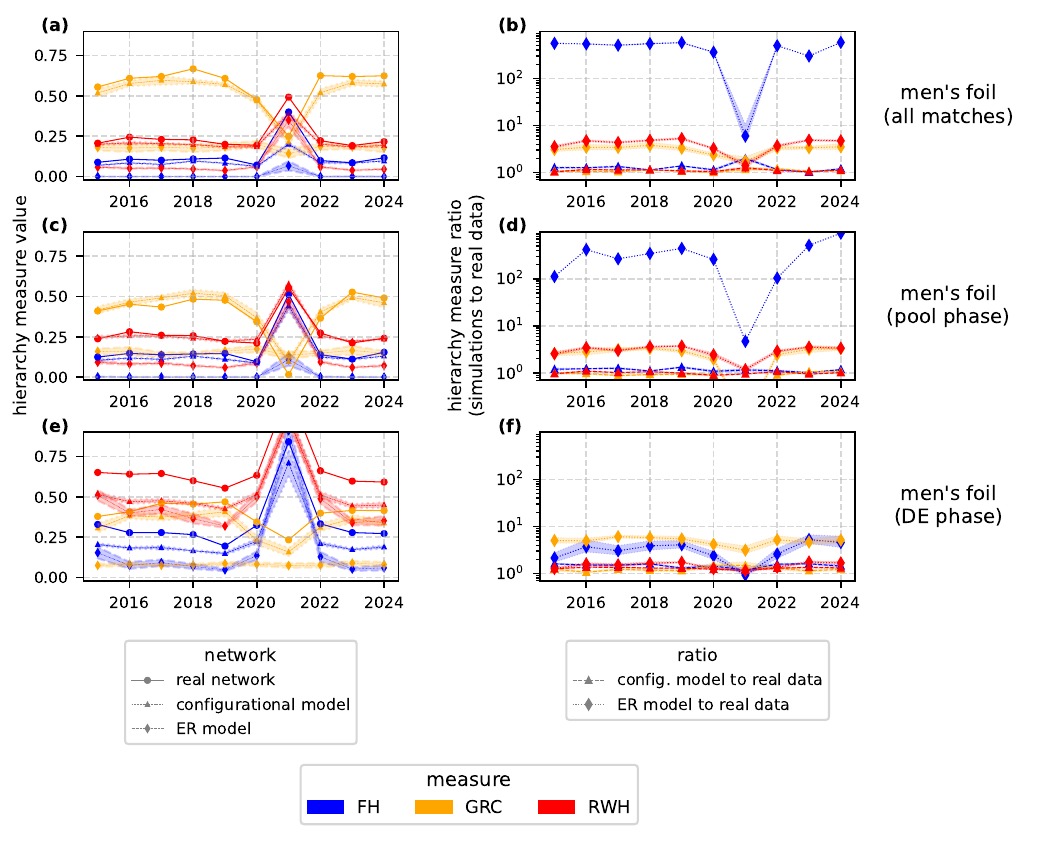}
    \caption{\textbf{Hierarchy measures in the network of men's foil matches.} \textbf{(a)} The flow hierarchy (blue), the global reaching centrality based on 2-reach (orange), and the random walk hierarchy (red) over time, displaying both the measured value in the original networks (circles) and also the average value in randomised networks according to the configuration model (triangles) and the Erdős--Rényi model (diamonds). \textbf{(b)} The three hierarchy values measured in the real network shown in panel (a), divided by the values in the randomised networks according to the configuration model (triangles) and the Erdős--Rényi model (diamonds).}
\end{figure}

\begin{figure}[h]
    \centering
    \includegraphics[width=6in]{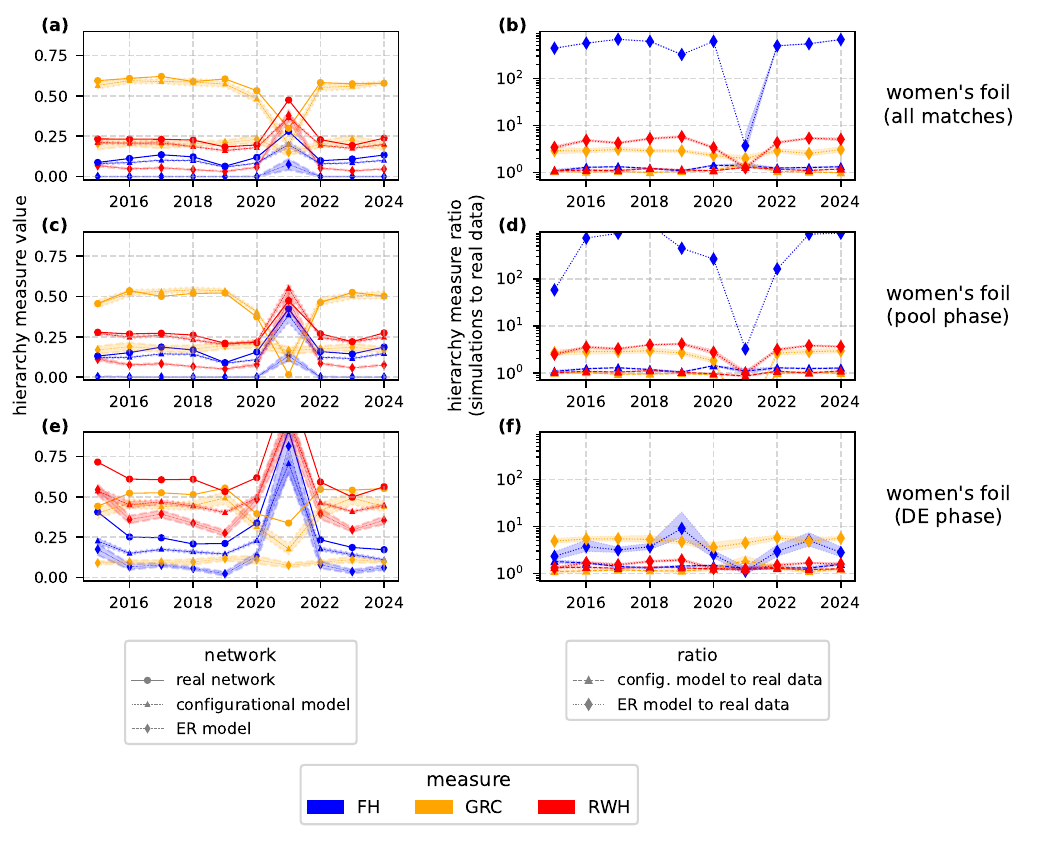}
    \caption{\textbf{Hierarchy measures in the network of women's foil matches.} \textbf{(a)} The flow hierarchy (blue), the global reaching centrality based on 2-reach (orange), and the random walk hierarchy (red) over time, displaying both the measured value in the original networks (circles) and also the average value in randomised networks according to the configuration model (triangles) and the Erdős--Rényi model (diamonds). \textbf{(b)} The three hierarchy values measured in the real network shown in panel (a), divided by the values in the randomised networks according to the configuration model (triangles) and the Erdős--Rényi model (diamonds).}
    \label{fig:hierarchy-2}
\end{figure}

\newpage \phantom{}
\newpage \phantom{}
\newpage \phantom{}
\newpage \phantom{}
\newpage \phantom{}
\newpage \phantom{}
\newpage \phantom{}
\newpage \phantom{}

\section*{3-cycle abundance in the network of matches}

\begin{figure}[h]
    \centering
    \includegraphics[width=6in]{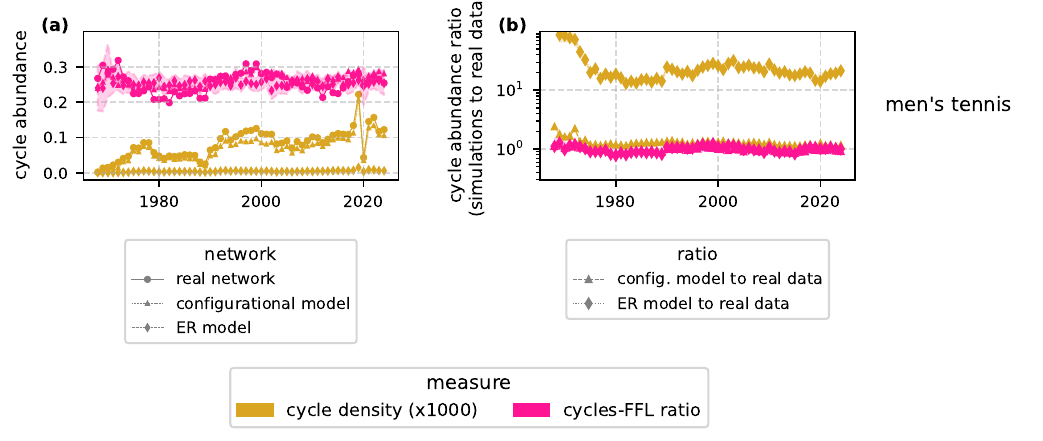}
    \caption{\textbf{3-cycle abundance in the network of men's tennis matches.} \textbf{(a)} The number of 3-cycles in the graph normed by the number all node triads (yellow), and its ratio to the number of feedforward 3-loops (FFLs) (pink) over time, displaying both the measured value in the original networks (circles), and also the average value in randomised networks according to the configuration model (triangles) and the Erdős--Rényi model (diamonds). The shaded regions around the averages indicate the standard deviation. \textbf{(b)} The two abundance values measured in the real network shown in panel (a), divided by the values in the randomised networks according to the configuration model (triangles) and the Erdős--Rényi model (diamonds).}
    \label{fig:cycles-1}
\end{figure}

In the right column of Figure 2 of the main text, we plotted two ways of measure the cycle abundance over time for the network of real matches, and the value of these measures for two types of network randomisation. These quantities measured how many cycles are there in the network relative to the size of the network. There, we presented the results of men's tennis tournaments, and men's sabre tournaments. In this section, we show the analogous results of all the studied dataset in Figures \ref{fig:cycles-1}-\ref{fig:cycles-2}.

\begin{figure}[h]
    \centering
    \includegraphics[width=6in]{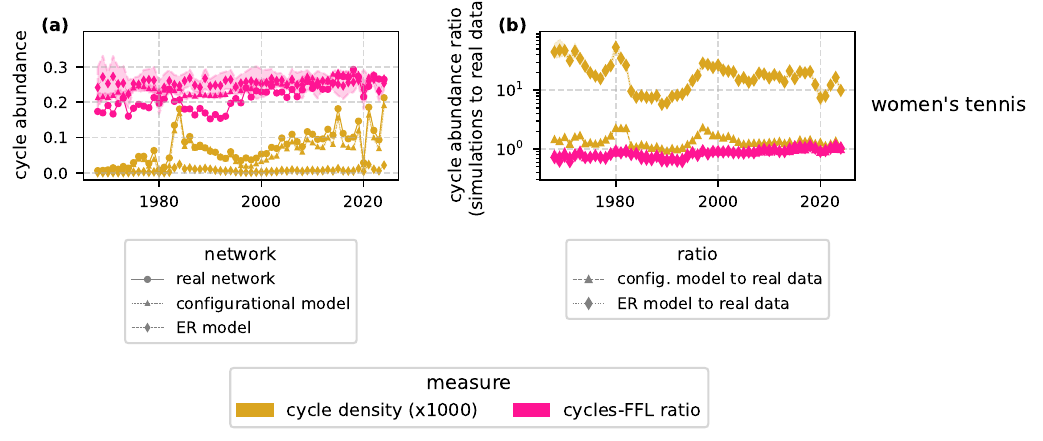}
    \caption{\textbf{3-cycle abundance in the network of women's tennis matches.} \textbf{(a)} The number of 3-cycles in the graph normed by the number all node triads (yellow), and its ratio to the number of feedforward 3-loops (FFLs) (pink) over time, displaying both the measured value in the original networks (circles), and also the average value in randomised networks according to the configuration model (triangles) and the Erdős--Rényi model (diamonds). The shaded regions around the averages indicate the standard deviation. \textbf{(b)} The two abundance values measured in the real network shown in panel (a), divided by the values in the randomised networks according to the configuration model (triangles) and the Erdős--Rényi model (diamonds).}
\end{figure}

\begin{figure}[h]
    \centering
    \includegraphics[width=6in]{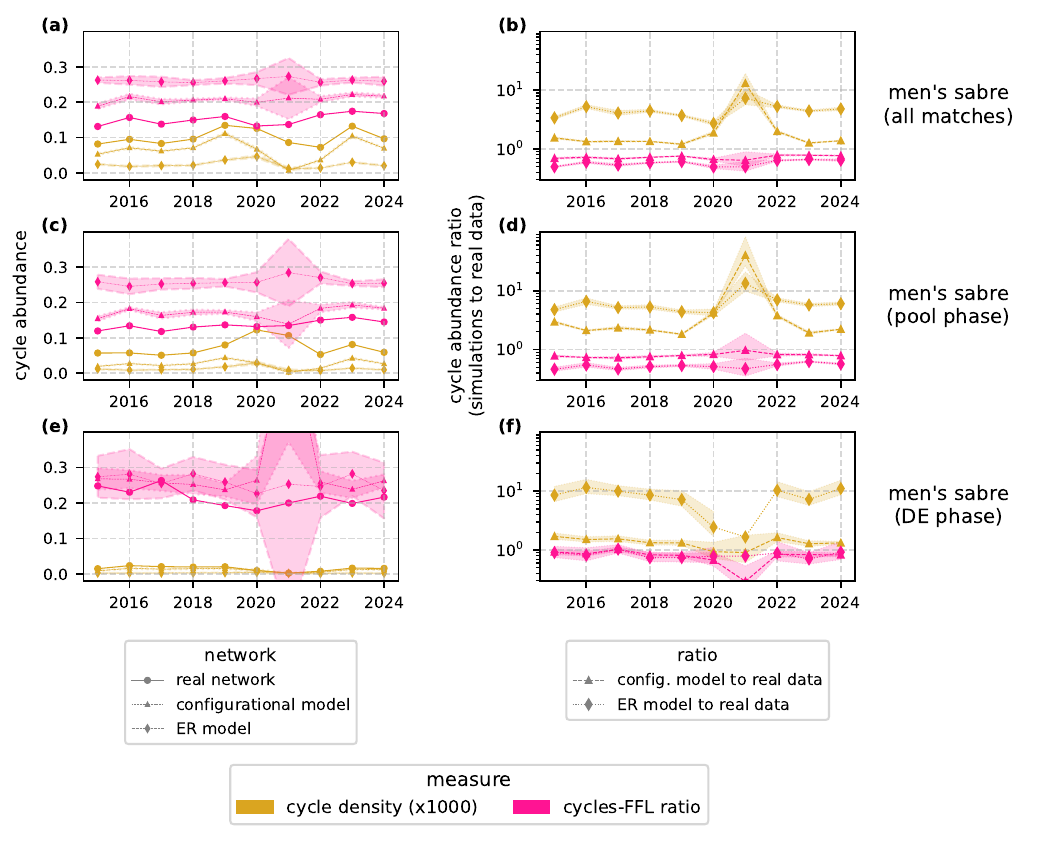}
    \caption{\textbf{3-cycle abundance in the network of men's sabre matches.} \textbf{(a)} The number of 3-cycles in the graph normed by the number all node triads (yellow), and its ratio to the number of feedforward 3-loops (FFLs) (pink) over time, displaying both the measured value in the original networks (circles), and also the average value in randomised networks according to the configuration model (triangles) and the Erdős--Rényi model (diamonds). The shaded regions around the averages indicate the standard deviation. \textbf{(b)} The two abundance values measured in the real network shown in panel (a), divided by the values in the randomised networks according to the configuration model (triangles) and the Erdős--Rényi model (diamonds).}
\end{figure}

\begin{figure}[h]
    \centering
    \includegraphics[width=6in]{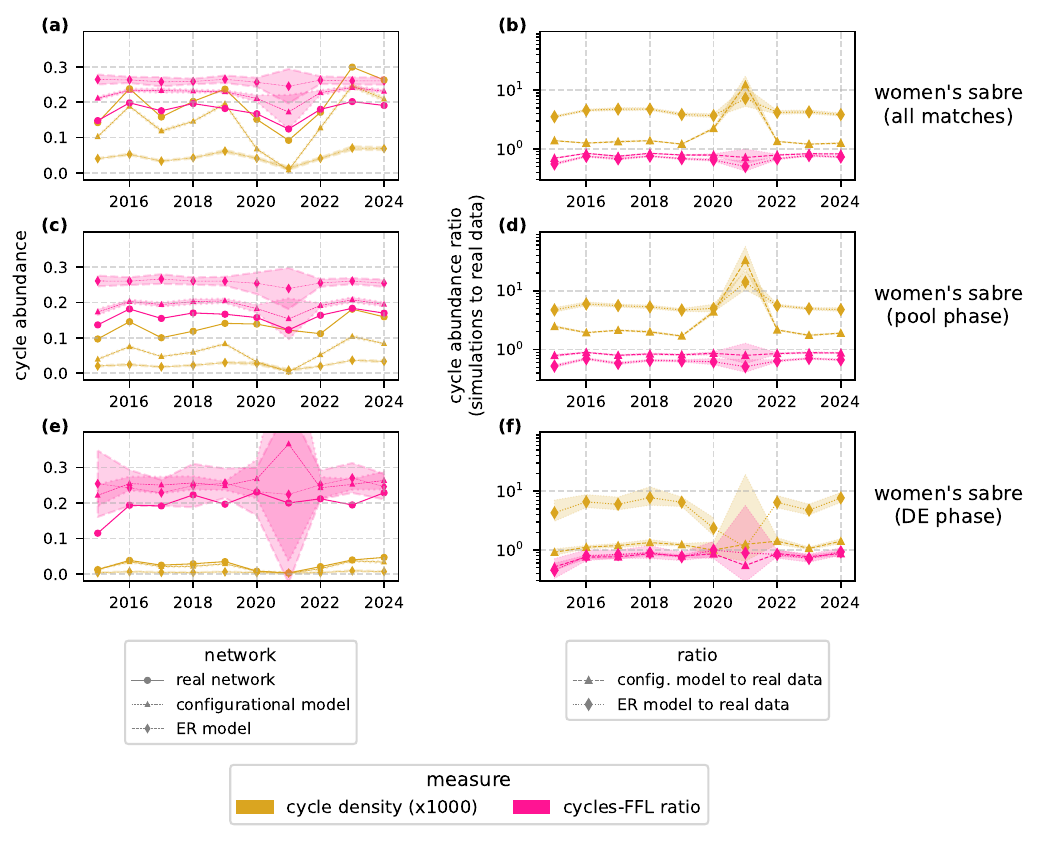}
    \caption{\textbf{3-cycle abundance in the network of women's sabre matches.} \textbf{(a)} The number of 3-cycles in the graph normed by the number all node triads (yellow), and its ratio to the number of feedforward 3-loops (FFLs) (pink) over time, displaying both the measured value in the original networks (circles), and also the average value in randomised networks according to the configuration model (triangles) and the Erdős--Rényi model (diamonds). The shaded regions around the averages indicate the standard deviation. \textbf{(b)} The two abundance values measured in the real network shown in panel (a), divided by the values in the randomised networks according to the configuration model (triangles) and the Erdős--Rényi model (diamonds).}
\end{figure}

\begin{figure}[h]
    \centering
    \includegraphics[width=6in]{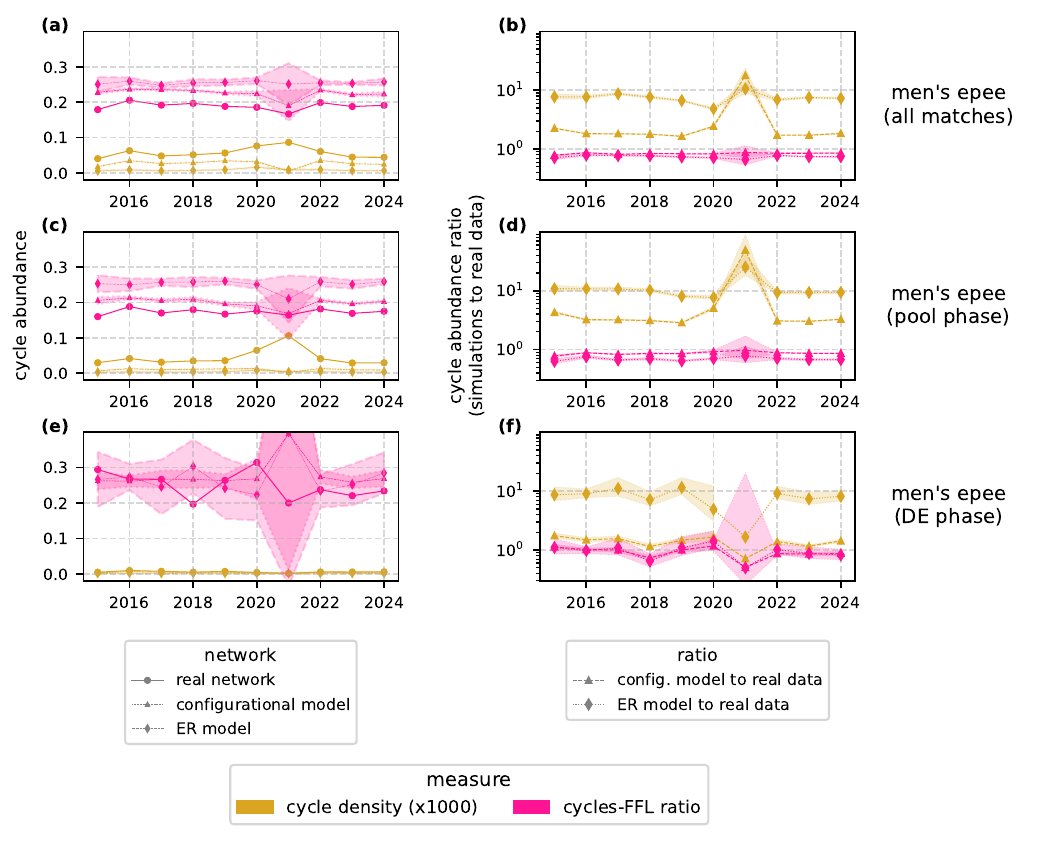}
    \caption{\textbf{3-cycle abundance in the network of men's épée matches.} \textbf{(a)} The number of 3-cycles in the graph normed by the number all node triads (yellow), and its ratio to the number of feedforward 3-loops (FFLs) (pink) over time, displaying both the measured value in the original networks (circles), and also the average value in randomised networks according to the configuration model (triangles) and the Erdős--Rényi model (diamonds). The shaded regions around the averages indicate the standard deviation. \textbf{(b)} The two abundance values measured in the real network shown in panel (a), divided by the values in the randomised networks according to the configuration model (triangles) and the Erdős--Rényi model (diamonds).}
\end{figure}

\begin{figure}[h]
    \centering
    \includegraphics[width=6in]{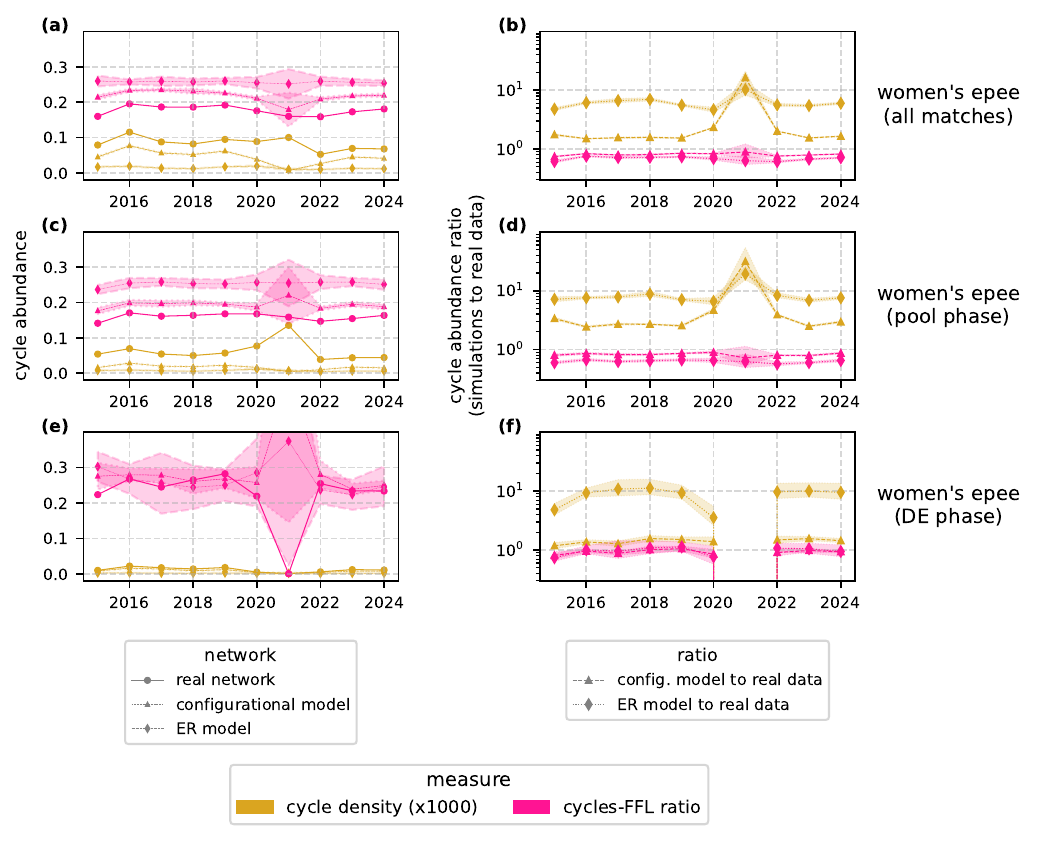}
    \caption{\textbf{3-cycle abundance in the network of women's épée matches.} \textbf{(a)} The number of 3-cycles in the graph normed by the number all node triads (yellow), and its ratio to the number of feedforward 3-loops (FFLs) (pink) over time, displaying both the measured value in the original networks (circles), and also the average value in randomised networks according to the configuration model (triangles) and the Erdős--Rényi model (diamonds). The shaded regions around the averages indicate the standard deviation. \textbf{(b)} The two abundance values measured in the real network shown in panel (a), divided by the values in the randomised networks according to the configuration model (triangles) and the Erdős--Rényi model (diamonds).}
\end{figure}

\begin{figure}[h]
    \centering
    \includegraphics[width=6in]{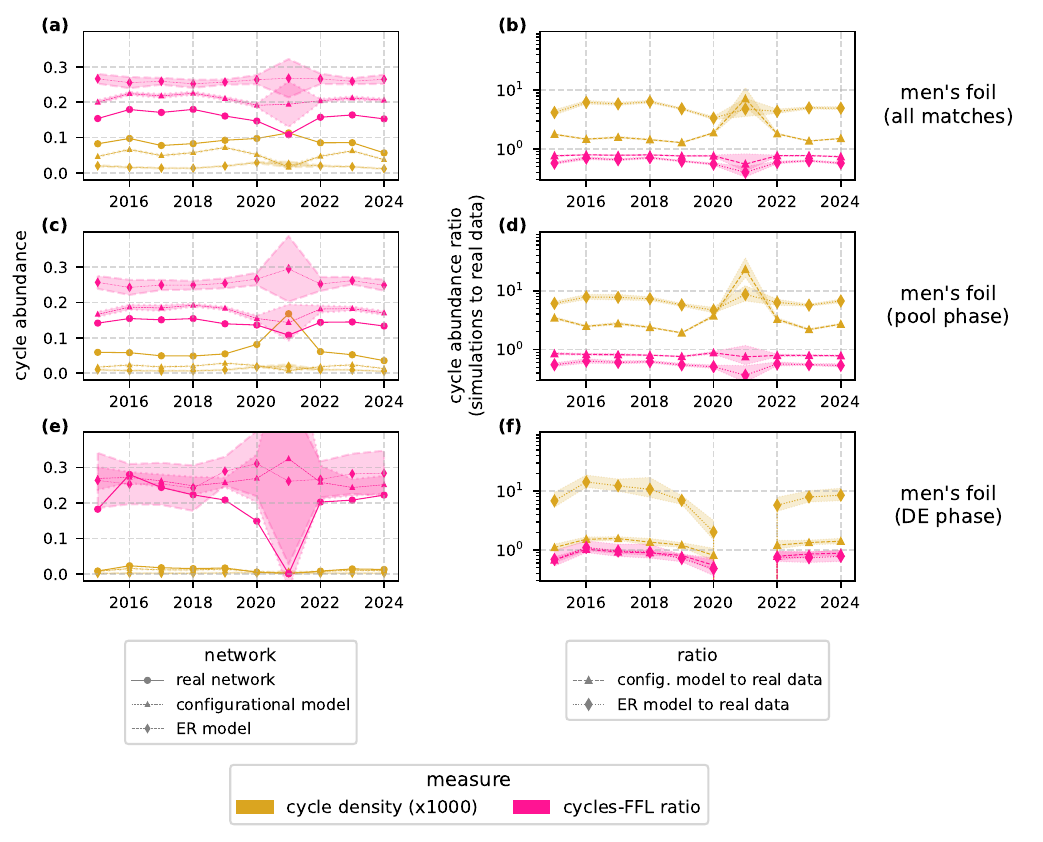}
    \caption{\textbf{3-cycle abundance in the network of men's foil matches.} \textbf{(a)} The number of 3-cycles in the graph normed by the number all node triads (yellow), and its ratio to the number of feedforward 3-loops (FFLs) (pink) over time, displaying both the measured value in the original networks (circles), and also the average value in randomised networks according to the configuration model (triangles) and the Erdős--Rényi model (diamonds). The shaded regions around the averages indicate the standard deviation. \textbf{(b)} The two abundance values measured in the real network shown in panel (a), divided by the values in the randomised networks according to the configuration model (triangles) and the Erdős--Rényi model (diamonds).}
\end{figure}

\begin{figure}[h]
    \centering
    \includegraphics[width=6in]{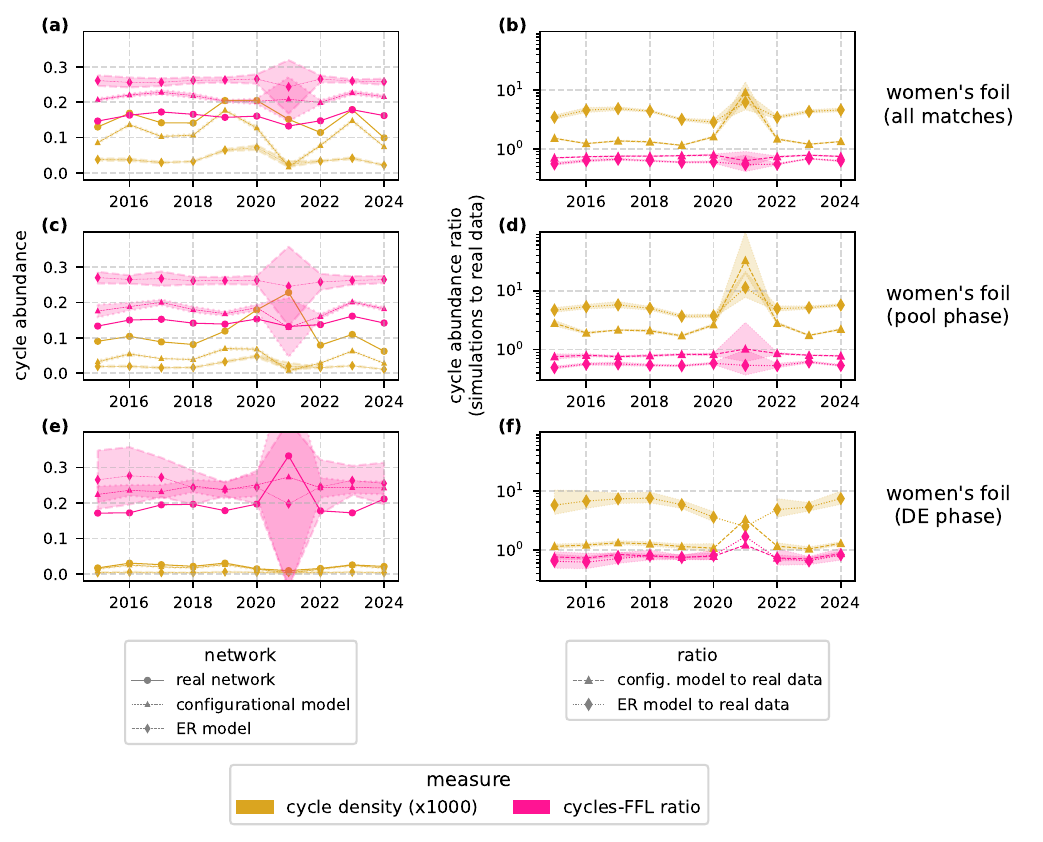}
    \caption{\textbf{3-cycle abundance in the network of women's foil matches.} \textbf{(a)} The number of 3-cycles in the graph normed by the number all node triads (yellow), and its ratio to the number of feedforward 3-loops (FFLs) (pink) over time, displaying both the measured value in the original networks (circles), and also the average value in randomised networks according to the configuration model (triangles) and the Erdős--Rényi model (diamonds). The shaded regions around the averages indicate the standard deviation. \textbf{(b)} The two abundance values measured in the real network shown in panel (a), divided by the values in the randomised networks according to the configuration model (triangles) and the Erdős--Rényi model (diamonds).}
    \label{fig:cycles-2}
\end{figure}

\newpage \phantom{}
\newpage \phantom{}
\newpage \phantom{}
\newpage \phantom{}
\newpage \phantom{}
\newpage \phantom{}
\newpage \phantom{}
\newpage \phantom{}

\section*{Cycle enrichment in the network of matches}

\begin{figure}[h]
    \centering
    \includegraphics[width=4.6in]{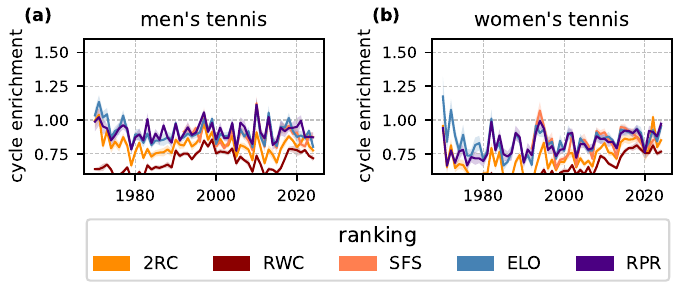}
    \caption{\textbf{Cycle enrichment calculated with all ranking scores in the network of matches.} The fraction of the number of 3-cycles and the expected number of 3-cycles from a given player strength distribution. The shaded regions around the averages indicate the standard deviation. \textbf{(a)} The cycle enrichment values for the network of men's tennis matches over the years. \textbf{(b)} The cycle enrichment values for the network of men's tennis matches over the years.}
    \label{fig:enrichment-1}
\end{figure}

In Figure 3 of the main text, we plotted the cycle enrichment calculated with SFS and ELO rankings over time for the network of matches. This quantity measured how much more 3-cycles are there in the network than expected based on the given ranking distribution. There, we presented the results using only the above two ranking scores, and only for men's tennis tournaments, and men's sabre tournaments. In this section, we show the analogous results of all the studied dataset in Figures \ref{fig:enrichment-1}-\ref{fig:enrichment-2}.

\begin{figure}[h]
    \centering
    \includegraphics[width=\textwidth]{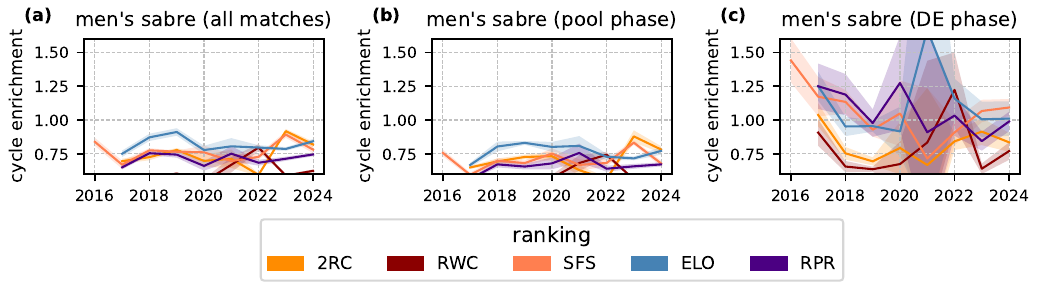}
    \caption{\textbf{Cycle enrichment calculated with all ranking scores in the network of matches.} The fraction of the number of 3-cycles and the expected number of 3-cycles from a given player strength distribution. The shaded regions around the averages indicate the standard deviation. \textbf{(a)} The cycle enrichment values for the network of men's sabre matches over the years. \textbf{(b)} The cycle enrichment values for the network of matches played in the pool phase in men's sabre tournaments. \textbf{(c)} The cycle enrichment values for the network of matches played in the direct elimination phase in men's sabre tournaments.}
\end{figure}

\begin{figure}[h]
    \centering
    \includegraphics[width=\textwidth]{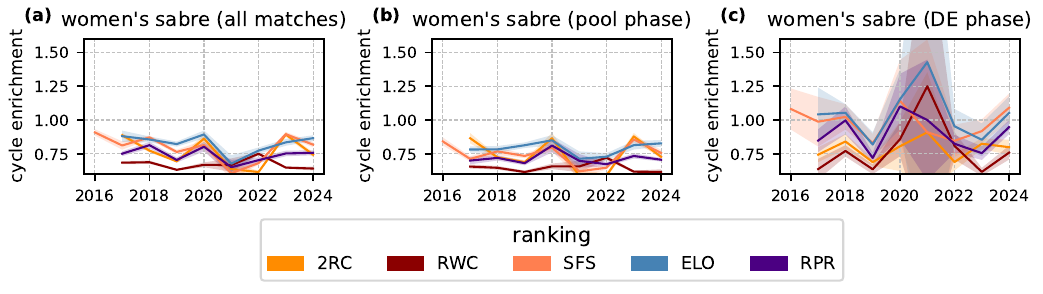}
    \caption{\textbf{Cycle enrichment calculated with all ranking scores in the network of matches.} The fraction of the number of 3-cycles and the expected number of 3-cycles from a given player strength distribution. The shaded regions around the averages indicate the standard deviation. \textbf{(a)} The cycle enrichment values for the network of women's sabre matches over the years. \textbf{(b)} The cycle enrichment values for the network of matches played in the pool phase in women's sabre tournaments. \textbf{(c)} The cycle enrichment values for the network of matches played in the direct elimination phase in women's sabre tournaments.}
\end{figure}

\begin{figure}[h]
    \centering
    \includegraphics[width=\textwidth]{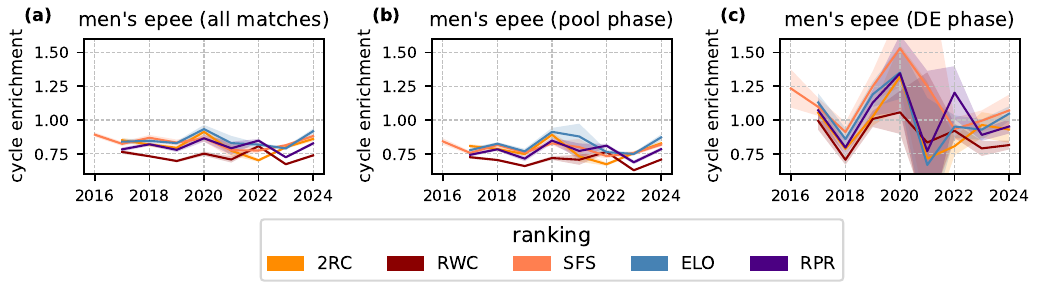}
    \caption{\textbf{Cycle enrichment calculated with all ranking scores in the network of matches.} The fraction of the number of 3-cycles and the expected number of 3-cycles from a given player strength distribution. The shaded regions around the averages indicate the standard deviation. \textbf{(a)} The cycle enrichment values for the network of men's épée matches over the years. \textbf{(b)} The cycle enrichment values for the network of matches played in the pool phase in men's épée tournaments. \textbf{(c)} The cycle enrichment values for the network of matches played in the direct elimination phase in men's épée tournaments.}
\end{figure}

\begin{figure}[h]
    \centering
    \includegraphics[width=\textwidth]{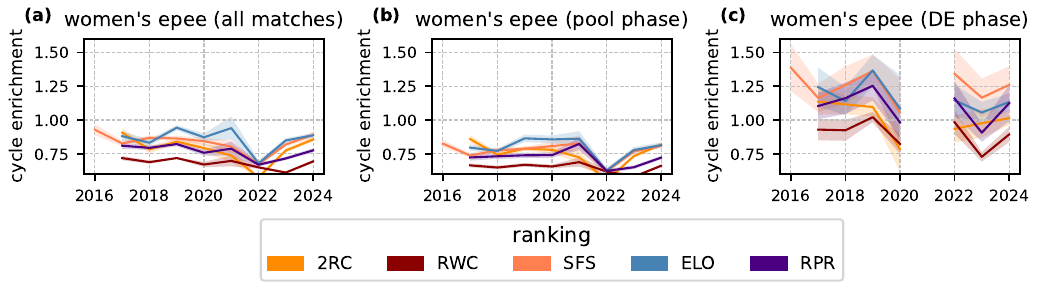}
    \caption{\textbf{Cycle enrichment calculated with all ranking scores in the network of matches.} The fraction of the number of 3-cycles and the expected number of 3-cycles from a given player strength distribution. The shaded regions around the averages indicate the standard deviation. \textbf{(a)} The cycle enrichment values for the network of women's épée matches over the years. \textbf{(b)} The cycle enrichment values for the network of matches played in the pool phase in women's épée tournaments. \textbf{(c)} The cycle enrichment values for the network of matches played in the direct elimination phase in women's épée tournaments.}
\end{figure}

\begin{figure}[h]
    \centering
    \includegraphics[width=\textwidth]{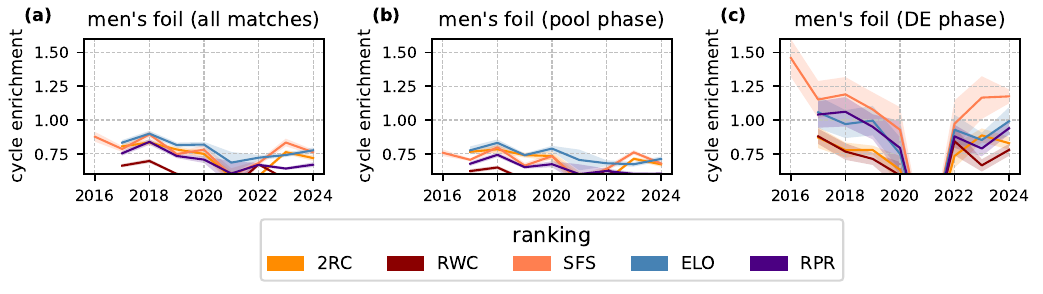}
    \caption{\textbf{Cycle enrichment calculated with all ranking scores in the network of matches.} The fraction of the number of 3-cycles and the expected number of 3-cycles from a given player strength distribution. The shaded regions around the averages indicate the standard deviation. \textbf{(a)} The cycle enrichment values for the network of men's foil matches over the years. \textbf{(b)} The cycle enrichment values for the network of matches played in the pool phase in men's foil tournaments. \textbf{(c)} The cycle enrichment values for the network of matches played in the direct elimination phase in men's foil tournaments.}
\end{figure}

\begin{figure}[h]
    \centering
    \includegraphics[width=\textwidth]{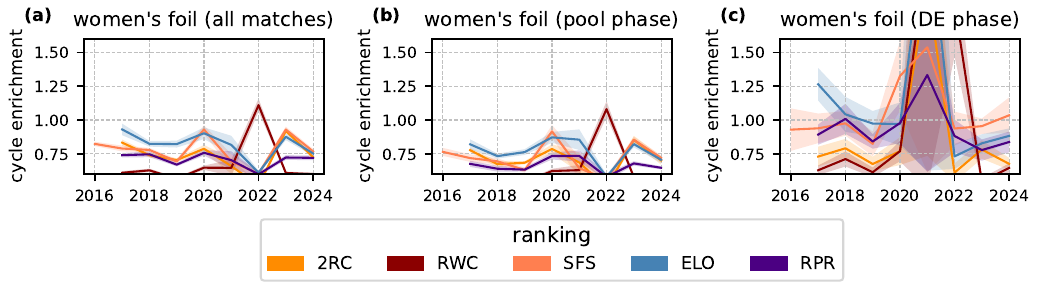}
    \caption{\textbf{Cycle enrichment calculated with all ranking scores in the network of matches.} The fraction of the number of 3-cycles and the expected number of 3-cycles from a given player strength distribution. The shaded regions around the averages indicate the standard deviation. \textbf{(a)} The cycle enrichment values for the network of women's foil matches over the years. \textbf{(b)} The cycle enrichment values for the network of matches played in the pool phase in women's foil tournaments. \textbf{(c)} The cycle enrichment values for the network of matches played in the direct elimination phase in women's foil tournaments.}
    \label{fig:enrichment-2}
\end{figure}
\newpage
\phantom{}
\newpage
\phantom{}
\newpage
\phantom{}
\newpage

\section*{The structure of the hierarchy}

\begin{figure}[h]
    \centering
    \includegraphics[width=\textwidth]{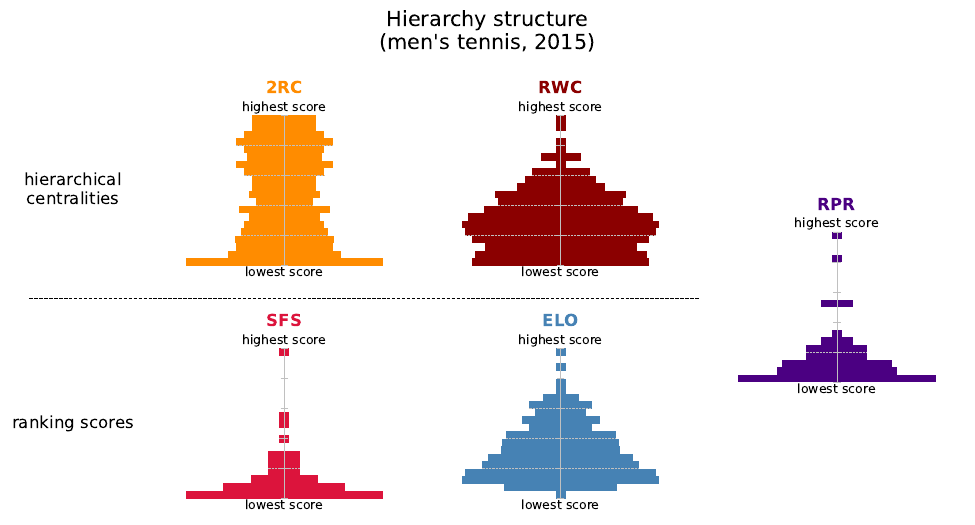}
    \caption{\textbf{The structure of the hierarchy for men's tennis in 2015.} The distribution of nodes along the 2-reach centrality (orange), the random walk centrality (red), the official sport ranking (pink), the Elo score (cyan), and the revered pagerank (purple) for the network of men's tennis matches played in 2015. The width of the bars represents the relative number of nodes having a specific centrality value on log-scale.}
    \label{fig:histogram-1}
\end{figure}

In Figures 4-5 of the main text, we presented the structure of the hierarchy in the network of men's tennis and men's sabre matches in 2015, using 5 different ranking scores. Such illustration can be done for the network of matches of any studied sport and from any year. In this section, we show the analogous results for all studied sports from the years 2015 and 2023 in Figures \ref{fig:histogram-1}-\ref{fig:histogram-2}.

\begin{figure}[h]
    \centering
    \includegraphics[width=\textwidth]{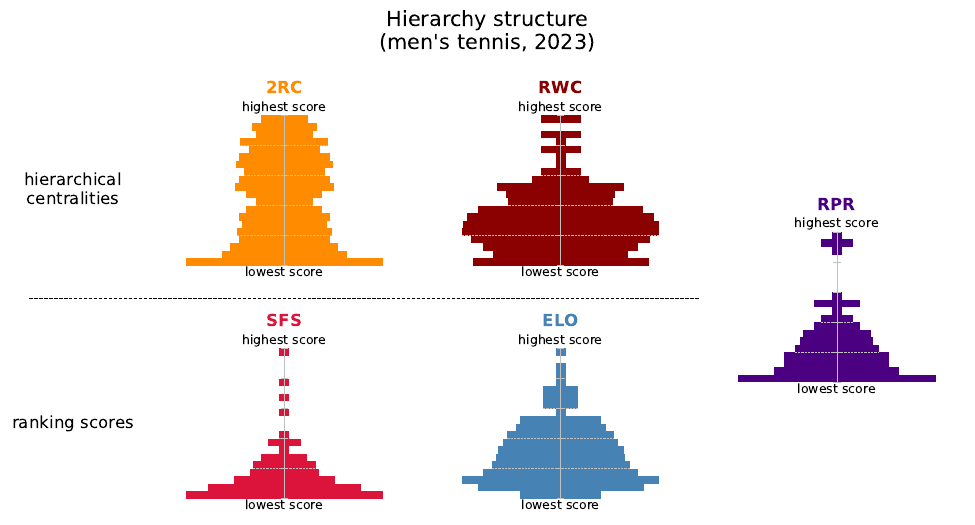}
    \caption{\textbf{The structure of the hierarchy for men's tennis in 2023.} The distribution of nodes along the 2-reach centrality (orange), the random walk centrality (red), the official sport ranking (pink), the Elo score (cyan), and the revered pagerank (purple) for the network of men's tennis matches played in 2023. The width of the bars represents the relative number of nodes having a specific centrality value on log-scale.}
\end{figure}

\newpage

\begin{figure}[p]
    \centering
    \includegraphics[width=\textwidth]{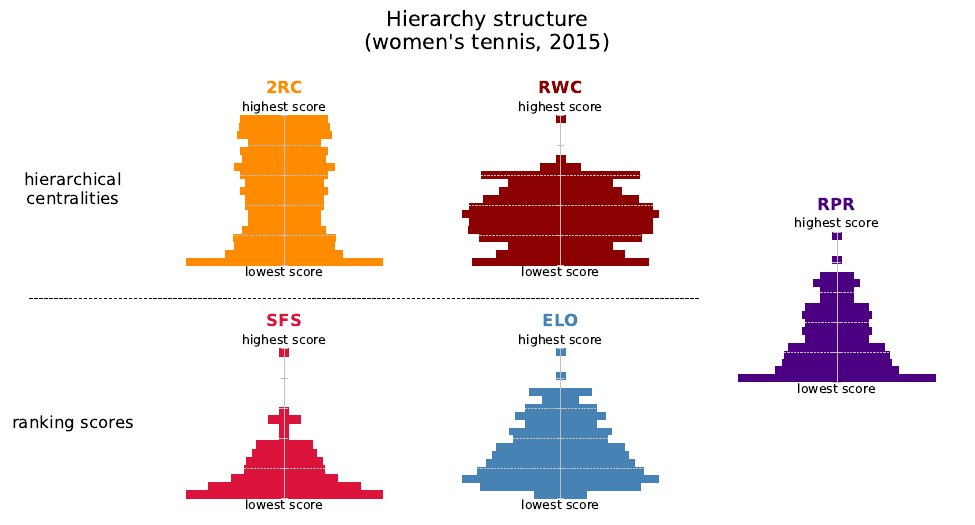}
    \caption{\textbf{The structure of the hierarchy for women's tennis in 2015.} The distribution of nodes along the 2-reach centrality (orange), the random walk centrality (red), the official sport ranking (pink), the Elo score (cyan), and the revered pagerank (purple) for the network of women's tennis matches played in 2015. The width of the bars represents the relative number of nodes having a specific centrality value on log-scale.}
\end{figure}

\begin{figure}[p]
    \centering
    \includegraphics[width=\textwidth]{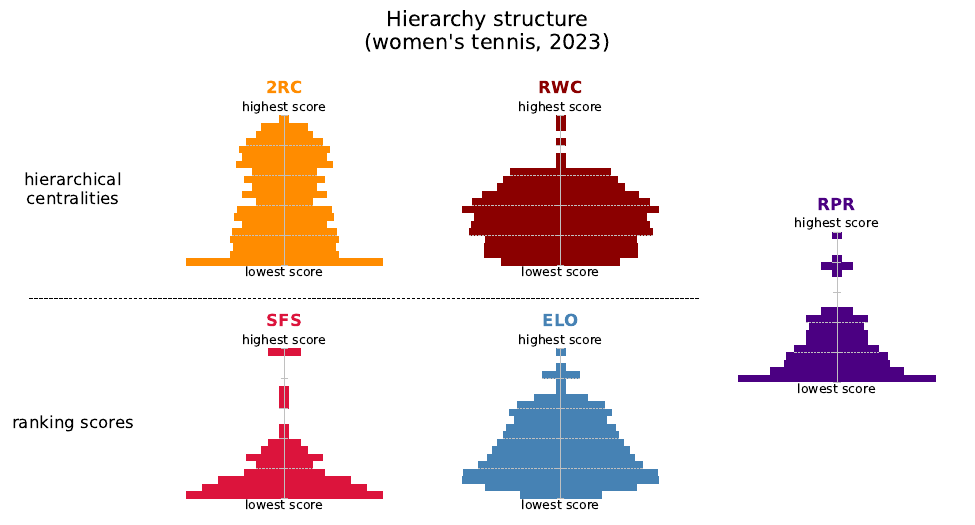}
    \caption{\textbf{The structure of the hierarchy for women's tennis in 2023.} The distribution of nodes along the 2-reach centrality (orange), the random walk centrality (red), the official sport ranking (pink), the Elo score (cyan), and the revered pagerank (purple) for the network of women's tennis matches played in 2023. The width of the bars represents the relative number of nodes having a specific centrality value on log-scale.}
\end{figure}

\begin{figure}[p]
    \centering
    \includegraphics[width=\textwidth]{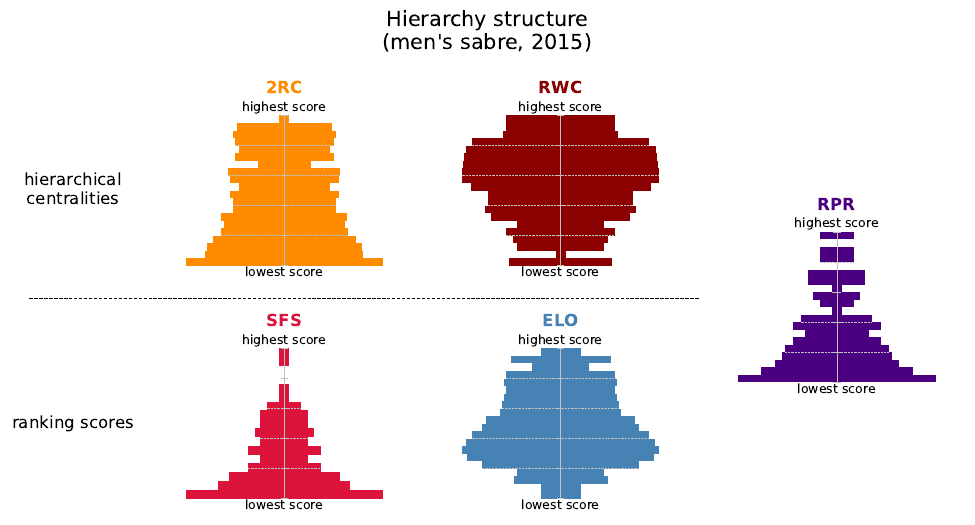}
    \caption{\textbf{The structure of the hierarchy for men's sabre in 2015.} The distribution of nodes along the 2-reach centrality (orange), the random walk centrality (red), the official sport ranking (pink), the Elo score (cyan), and the revered pagerank (purple) for the network of men's sabre matches played in 2015. The width of the bars represents the relative number of nodes having a specific centrality value on log-scale.}
\end{figure}

\begin{figure}[p]
    \centering
    \includegraphics[width=\textwidth]{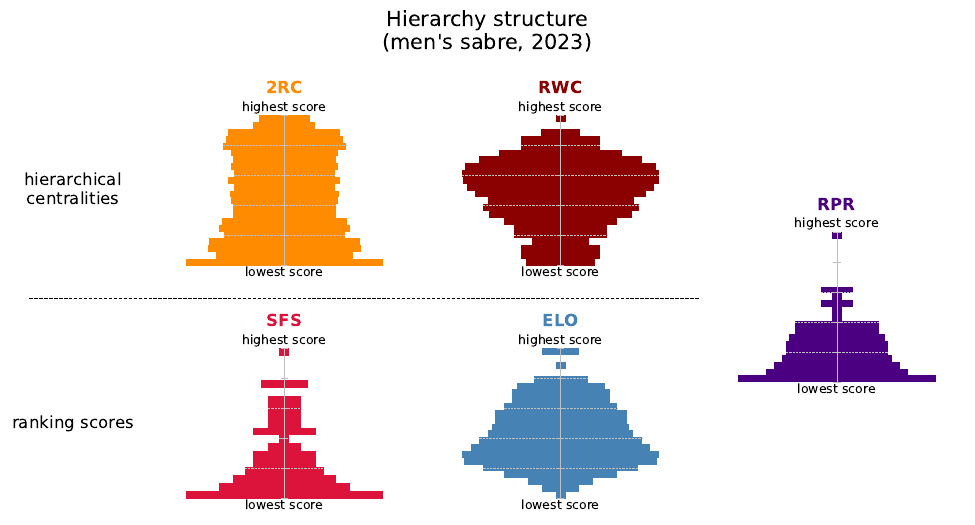}
    \caption{\textbf{The structure of the hierarchy for men's sabre in 2023.} The distribution of nodes along the 2-reach centrality (orange), the random walk centrality (red), the official sport ranking (pink), the Elo score (cyan), and the revered pagerank (purple) for the network of men's sabre matches played in 2023. The width of the bars represents the relative number of nodes having a specific centrality value on log-scale.}
\end{figure}

\begin{figure}[p]
    \centering
    \includegraphics[width=\textwidth]{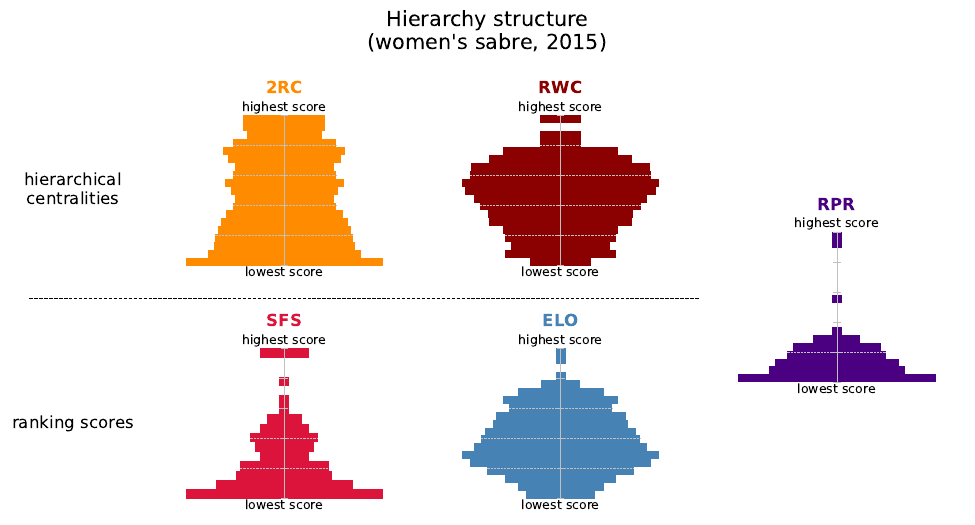}
    \caption{\textbf{The structure of the hierarchy for women's sabre in 2015.} The distribution of nodes along the 2-reach centrality (orange), the random walk centrality (red), the official sport ranking (pink), the Elo score (cyan), and the revered pagerank (purple) for the network of women's sabre matches played in 2015. The width of the bars represents the relative number of nodes having a specific centrality value on log-scale.}
\end{figure}

\begin{figure}[p]
    \centering
    \includegraphics[width=\textwidth]{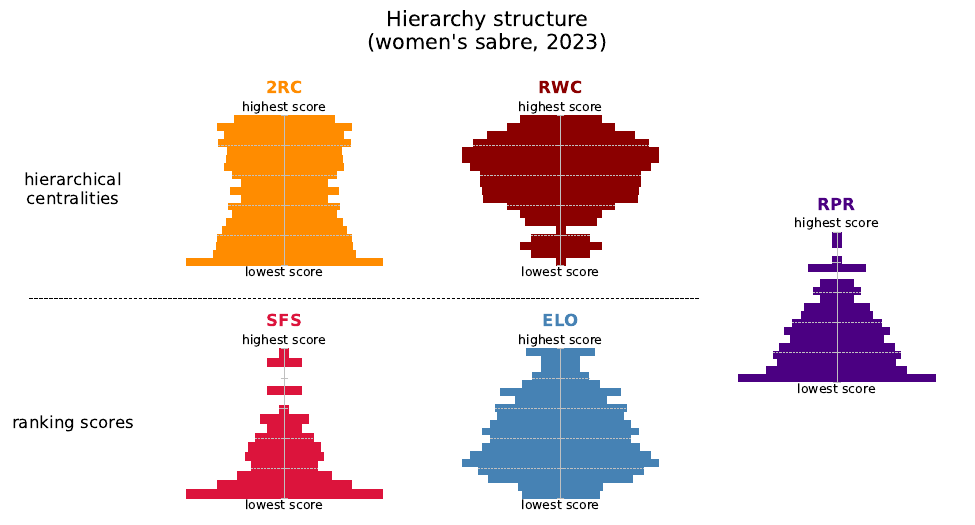}
    \caption{\textbf{The structure of the hierarchy for women's sabre in 2023.} The distribution of nodes along the 2-reach centrality (orange), the random walk centrality (red), the official sport ranking (pink), the Elo score (cyan), and the revered pagerank (purple) for the network of women's sabre matches played in 2023. The width of the bars represents the relative number of nodes having a specific centrality value on log-scale.}
\end{figure}

\begin{figure}[p]
    \centering
    \includegraphics[width=\textwidth]{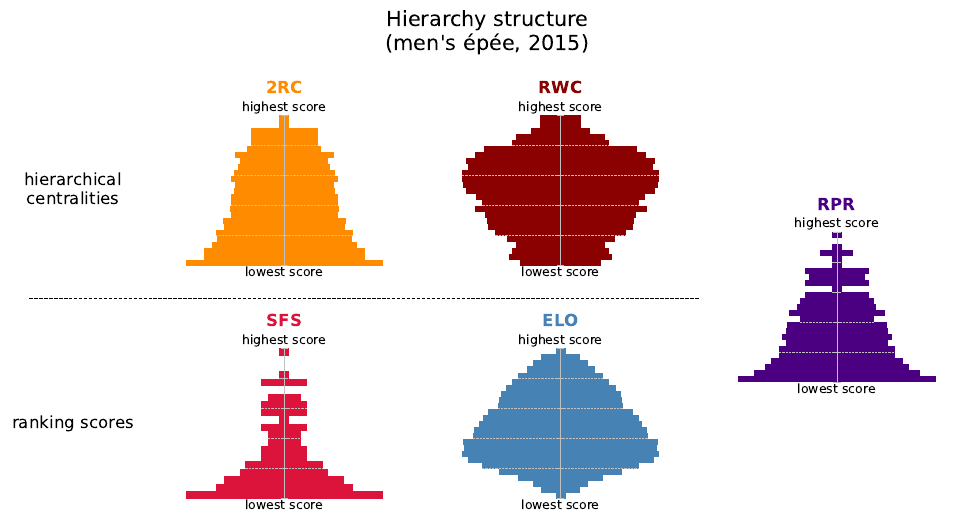}
    \caption{\textbf{The structure of the hierarchy for men's épée in 2015.} The distribution of nodes along the 2-reach centrality (orange), the random walk centrality (red), the official sport ranking (pink), the Elo score (cyan), and the revered pagerank (purple) for the network of men's épée matches played in 2015. The width of the bars represents the relative number of nodes having a specific centrality value on log-scale.}
\end{figure}

\begin{figure}[p]
    \centering
    \includegraphics[width=\textwidth]{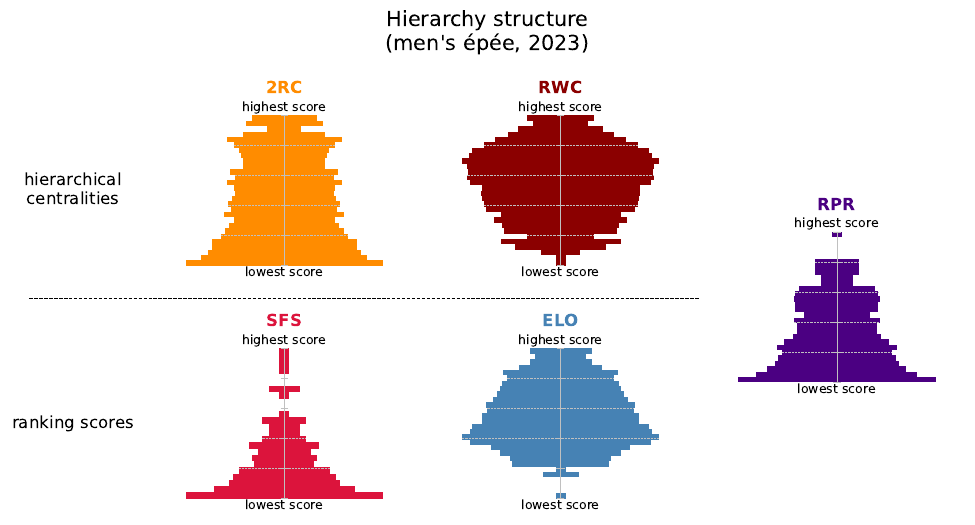}
    \caption{\textbf{The structure of the hierarchy for men's épée in 2023.} The distribution of nodes along the 2-reach centrality (orange), the random walk centrality (red), the official sport ranking (pink), the Elo score (cyan), and the revered pagerank (purple) for the network of men's épée matches played in 2023. The width of the bars represents the relative number of nodes having a specific centrality value on log-scale.}
\end{figure}

\begin{figure}[p]
    \centering
    \includegraphics[width=\textwidth]{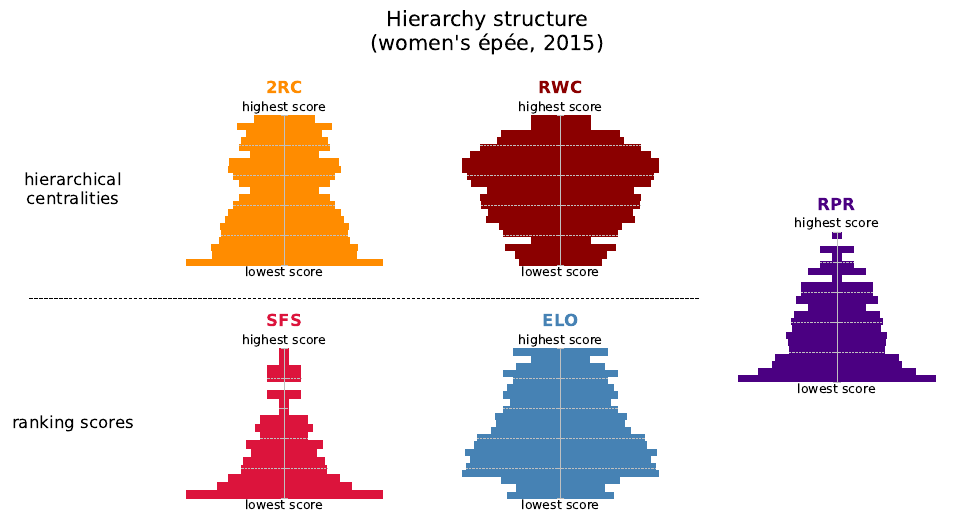}
    \caption{\textbf{The structure of the hierarchy for women's épée in 2015.} The distribution of nodes along the 2-reach centrality (orange), the random walk centrality (red), the official sport ranking (pink), the Elo score (cyan), and the revered pagerank (purple) for the network of women's épée matches played in 2015. The width of the bars represents the relative number of nodes having a specific centrality value on log-scale.}
\end{figure}

\begin{figure}[p]
    \centering
    \includegraphics[width=\textwidth]{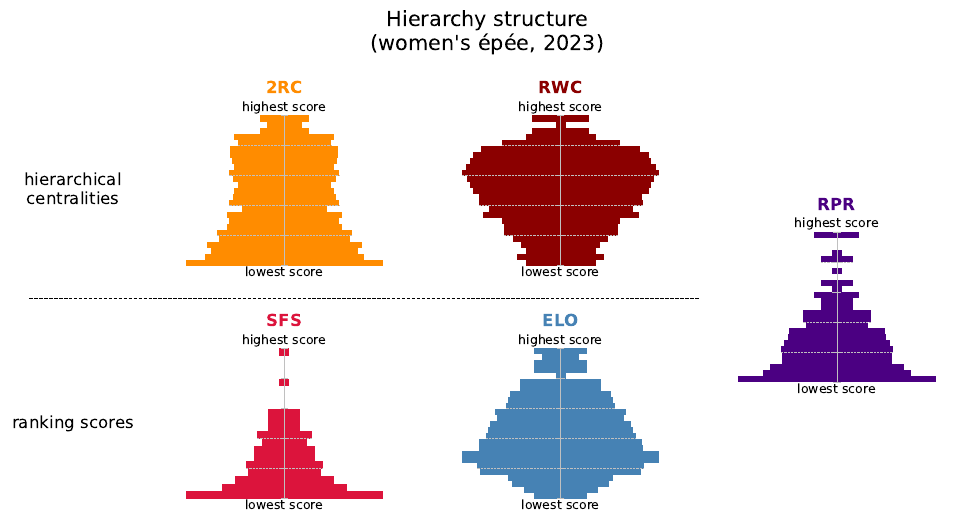}
    \caption{\textbf{The structure of the hierarchy for women's épée in 2023.} The distribution of nodes along the 2-reach centrality (orange), the random walk centrality (red), the official sport ranking (pink), the Elo score (cyan), and the revered pagerank (purple) for the network of women's épée matches played in 2023. The width of the bars represents the relative number of nodes having a specific centrality value on log-scale.}
\end{figure}

\begin{figure}[p]
    \centering
    \includegraphics[width=\textwidth]{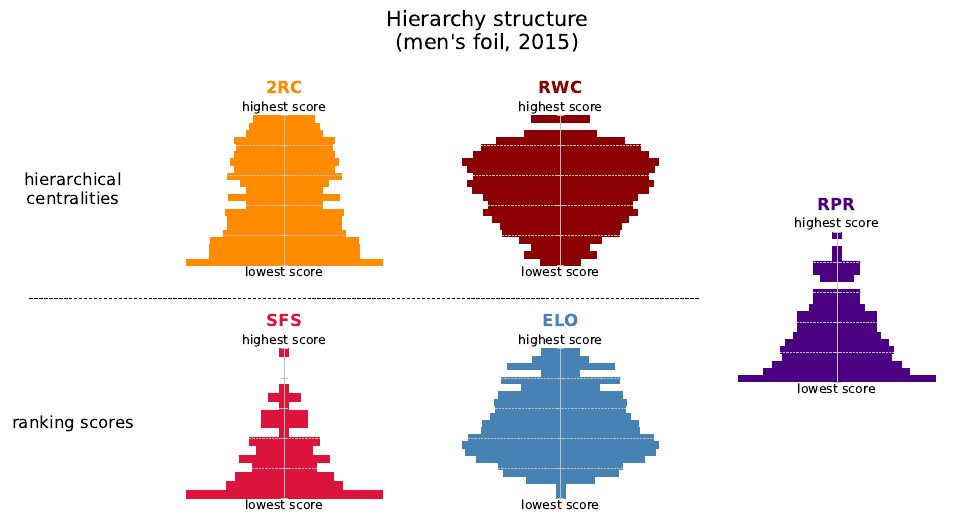}
    \caption{\textbf{The structure of the hierarchy for men's foil in 2015.} The distribution of nodes along the 2-reach centrality (orange), the random walk centrality (red), the official sport ranking (pink), the Elo score (cyan), and the revered pagerank (purple) for the network of men's foil matches played in 2015. The width of the bars represents the relative number of nodes having a specific centrality value on log-scale.}
\end{figure}

\begin{figure}[p]
    \centering
    \includegraphics[width=\textwidth]{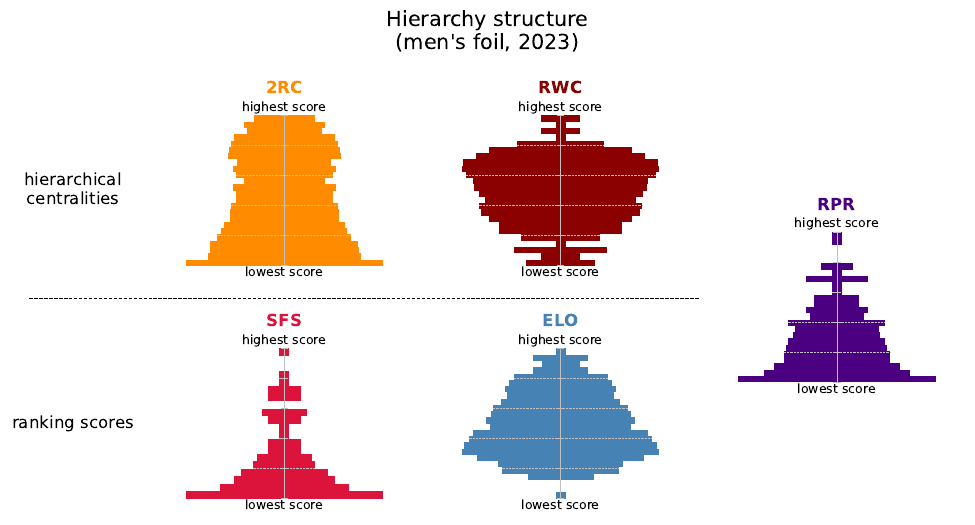}
    \caption{\textbf{The structure of the hierarchy for men's foil in 2023.} The distribution of nodes along the 2-reach centrality (orange), the random walk centrality (red), the official sport ranking (pink), the Elo score (cyan), and the revered pagerank (purple) for the network of men's foil matches played in 2023. The width of the bars represents the relative number of nodes having a specific centrality value on log-scale.}
\end{figure}

\begin{figure}[p]
    \centering
    \includegraphics[width=\textwidth]{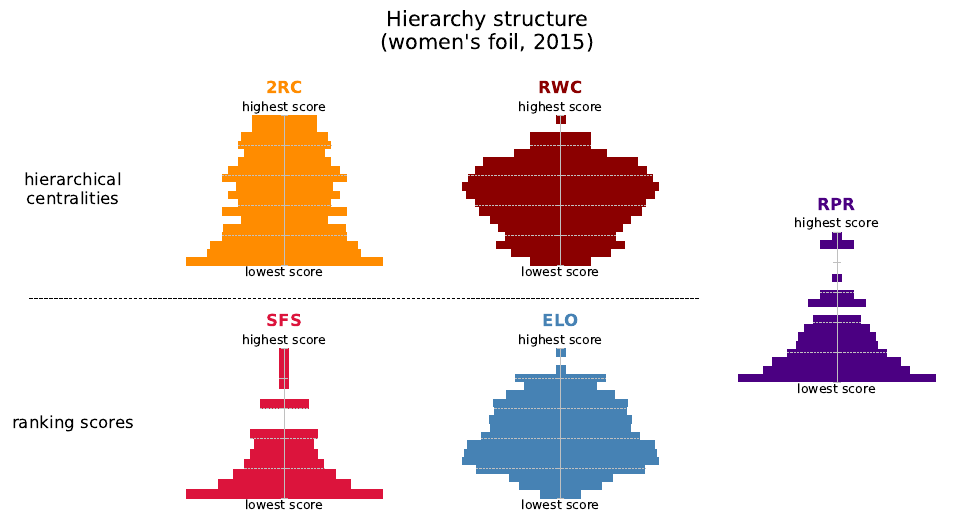}
    \caption{\textbf{The structure of the hierarchy for women's foil in 2015.} The distribution of nodes along the 2-reach centrality (orange), the random walk centrality (red), the official sport ranking (pink), the Elo score (cyan), and the revered pagerank (purple) for the network of women's foil matches played in 2015. The width of the bars represents the relative number of nodes having a specific centrality value on log-scale.}
\end{figure}

\begin{figure}[p]
    \centering
    \includegraphics[width=\textwidth]{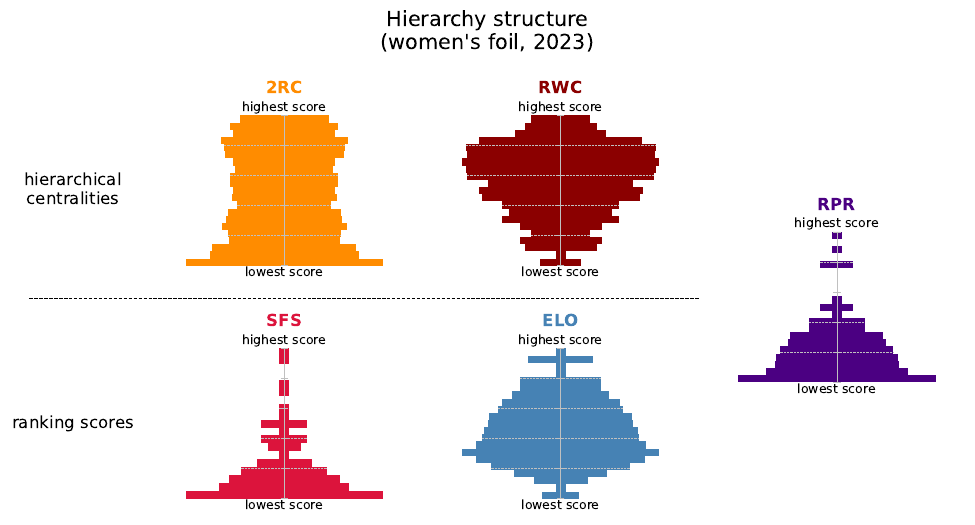}
    \caption{\textbf{The structure of the hierarchy for women's foil in 2023.} The distribution of nodes along the 2-reach centrality (orange), the random walk centrality (red), the official sport ranking (pink), the Elo score (cyan), and the revered pagerank (purple) for the network of women's foil matches played in 2023. The width of the bars represents the relative number of nodes having a specific centrality value on log-scale.}
    \label{fig:histogram-2}
\end{figure}

\newpage
\phantom{}
\newpage
\phantom{}

\section*{The rank--score distribution of local measures}

\begin{figure}[h]
    \centering
    \includegraphics[width=\textwidth]{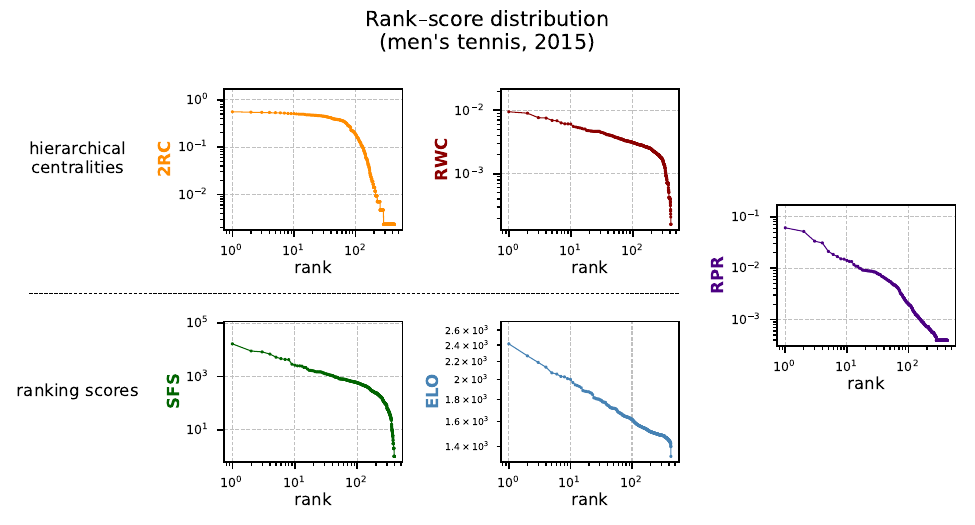}
    \caption{\textbf{The rank--score distribution of local measures for men's tennis in 2015.} The rank--score distribution of nodes along the 2-reach centrality (orange), the random walk centrality (red), the official sport ranking (pink), the Elo score (cyan), and the reverse pagerank (purple) for the network of men's tennis matches played in 2015. In the case of reversed pagerank two segments can be found following power-law scaling.}
    \label{fig:zipf-1}
\end{figure}

In Figures 6-7 of the main text, we presented the rank--socre distribution of athletes in men's tennis and men's sabre in 2015, using 5 different ranking scores. Such illustration can be done for the network of matches of any studied sport and from any year. In this section, we show the analogous results for all studied sports from the years 2015 and 2023 in Figures \ref{fig:zipf-1}-\ref{fig:zipf-2}.

\begin{figure}[h]
    \centering
    \includegraphics[width=\textwidth]{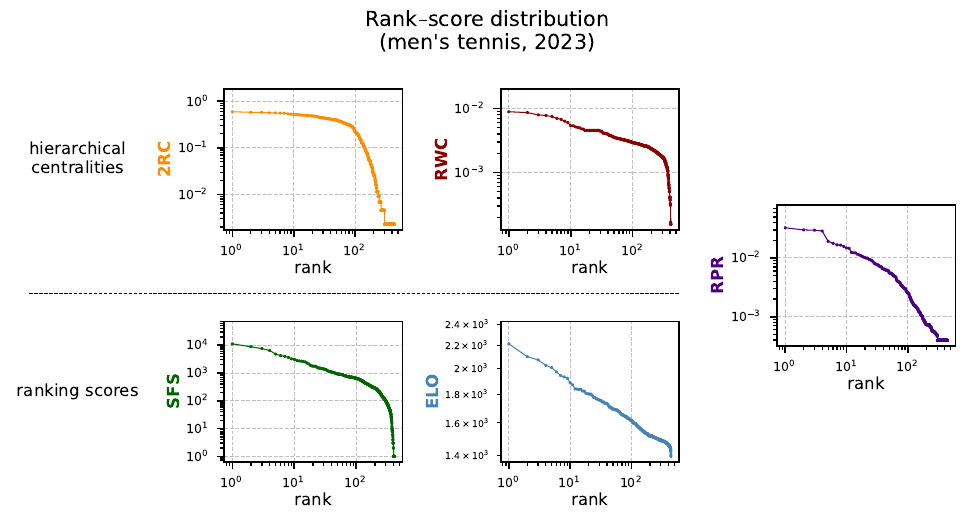}
    \caption{\textbf{The rank--score distribution of local measures for men's tennis in 2015.} The rank--score distribution of nodes along the 2-reach centrality (orange), the random walk centrality (red), the official sport ranking (pink), the Elo score (cyan), and the reverse pagerank (purple) for the network of men's tennis matches played in 2015. In the case of reversed pagerank two segments can be found following power-law scaling.}
\end{figure}

\newpage
\phantom{}

\begin{figure}[p]
    \centering
    \includegraphics[width=\textwidth]{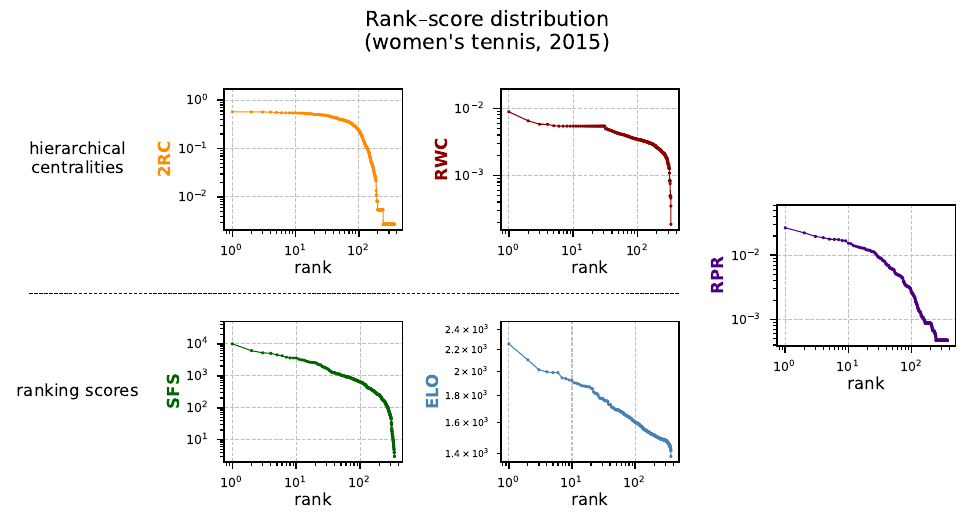}
    \caption{\textbf{The rank--score distribution of local measures for women's tennis in 2015.} The rank--score distribution of nodes along the 2-reach centrality (orange), the random walk centrality (red), the official sport ranking (pink), the Elo score (cyan), and the reverse pagerank (purple) for the network of women's tennis matches played in 2015. In the case of reversed pagerank two segments can be found following power-law scaling.}
\end{figure}

\begin{figure}[p]
    \centering
    \includegraphics[width=\textwidth]{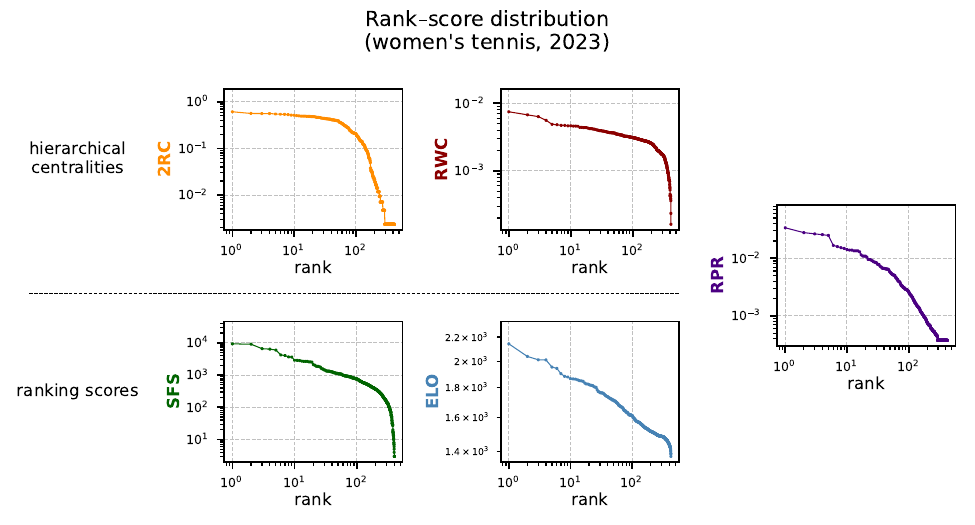}
    \caption{\textbf{The rank--score distribution of local measures for women's tennis in 2023.} The rank--score distribution of nodes along the 2-reach centrality (orange), the random walk centrality (red), the official sport ranking (pink), the Elo score (cyan), and the reverse pagerank (purple) for the network of women's tennis matches played in 2023. In the case of reversed pagerank two segments can be found following power-law scaling.}
\end{figure}

\begin{figure}[p]
    \centering
    \includegraphics[width=\textwidth]{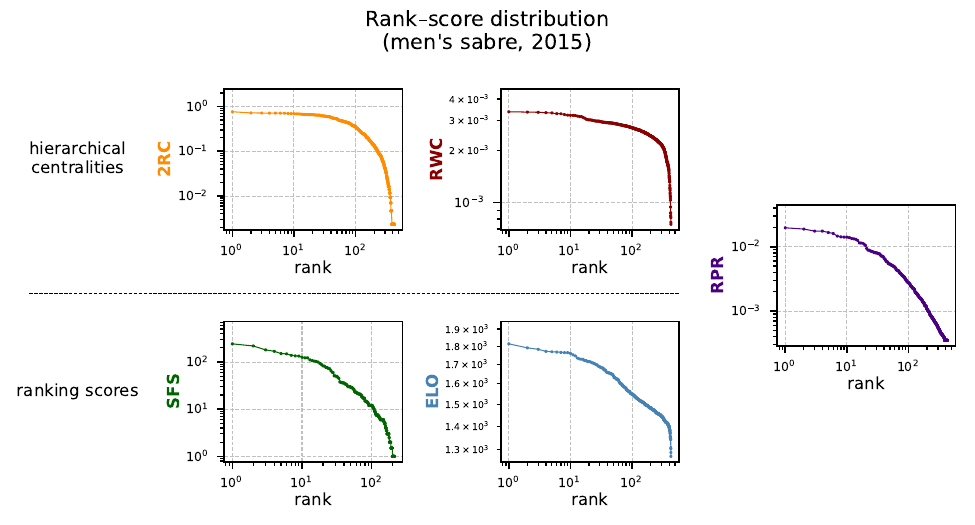}
    \caption{\textbf{The rank--score distribution of local measures for men's sabre in 2015.} The rank--score distribution of nodes along the 2-reach centrality (orange), the random walk centrality (red), the official sport ranking (pink), the Elo score (cyan), and the reverse pagerank (purple) for the network of men's sabre matches played in 2015. In the case of reversed pagerank two segments can be found following power-law scaling.}
\end{figure}

\begin{figure}[p]
    \centering
    \includegraphics[width=\textwidth]{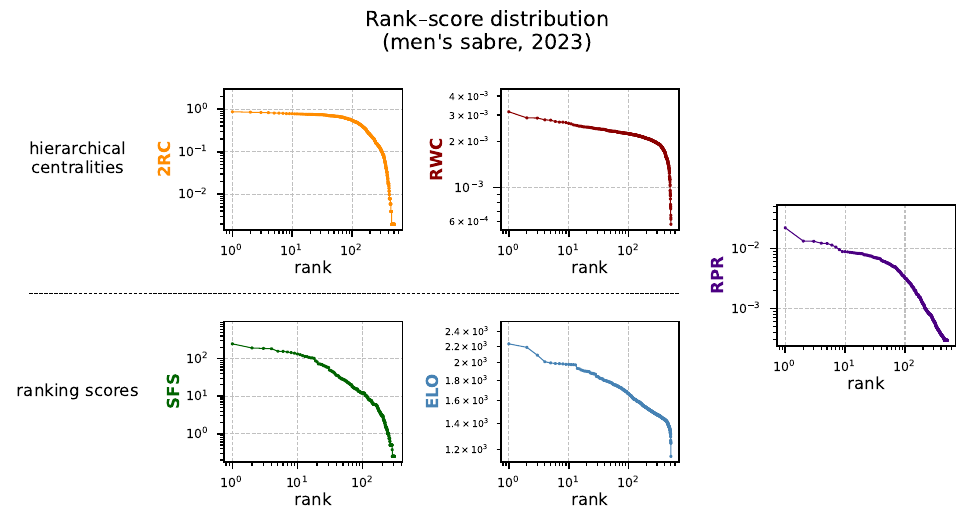}
    \caption{\textbf{The rank--score distribution of local measures for men's sabre in 2023.} The rank--score distribution of nodes along the 2-reach centrality (orange), the random walk centrality (red), the official sport ranking (pink), the Elo score (cyan), and the reverse pagerank (purple) for the network of men's sabre matches played in 2023. In the case of reversed pagerank two segments can be found following power-law scaling.}
\end{figure}

\begin{figure}[p]
    \centering
    \includegraphics[width=\textwidth]{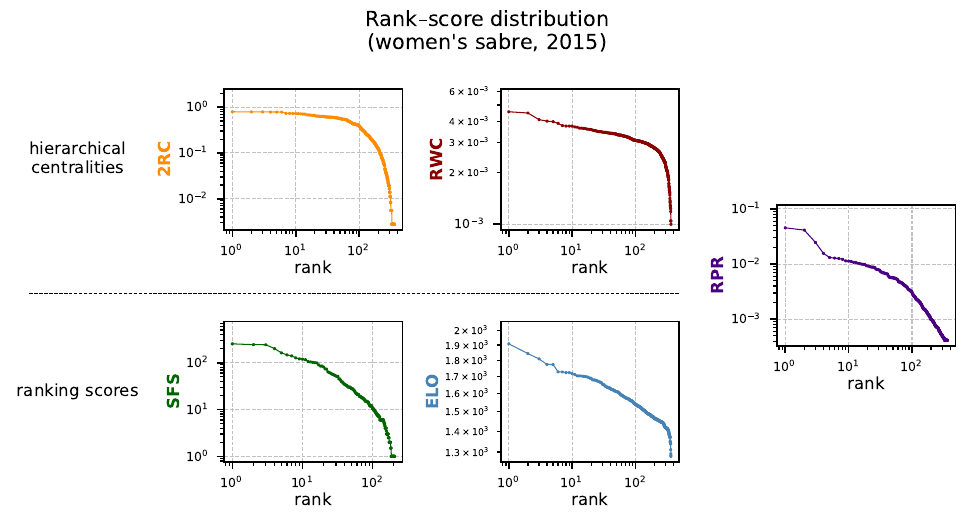}
    \caption{\textbf{The rank--score distribution of local measures for women's sabre in 2015.} The rank--score distribution of nodes along the 2-reach centrality (orange), the random walk centrality (red), the official sport ranking (pink), the Elo score (cyan), and the reverse pagerank (purple) for the network of women's sabre matches played in 2015. In the case of reversed pagerank two segments can be found following power-law scaling.}
\end{figure}

\begin{figure}[p]
    \centering
    \includegraphics[width=\textwidth]{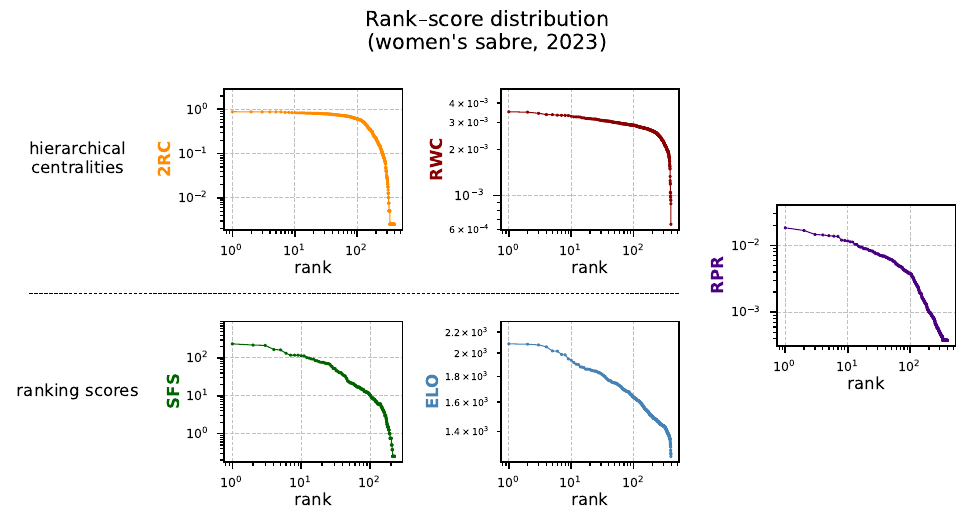}
    \caption{\textbf{The rank--score distribution of local measures for women's sabre in 2023.} The rank--score distribution of nodes along the 2-reach centrality (orange), the random walk centrality (red), the official sport ranking (pink), the Elo score (cyan), and the reverse pagerank (purple) for the network of women's sabre matches played in 2023. In the case of reversed pagerank two segments can be found following power-law scaling.}
\end{figure}

\begin{figure}[p]
    \centering
    \includegraphics[width=\textwidth]{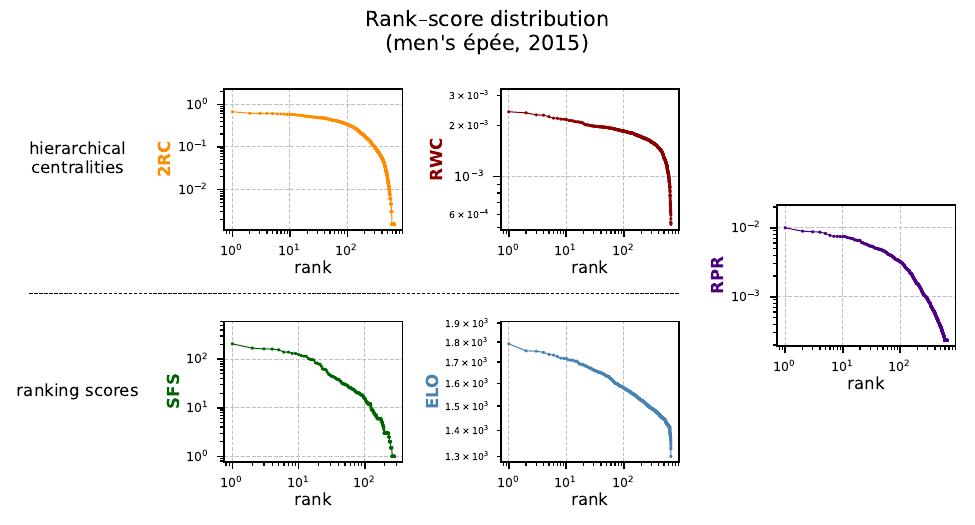}
    \caption{\textbf{The rank--score distribution of local measures for men's épée in 2015.} The rank--score distribution of nodes along the 2-reach centrality (orange), the random walk centrality (red), the official sport ranking (pink), the Elo score (cyan), and the reverse pagerank (purple) for the network of men's épée matches played in 2015. In the case of reversed pagerank two segments can be found following power-law scaling.}
\end{figure}

\begin{figure}[p]
    \centering
    \includegraphics[width=\textwidth]{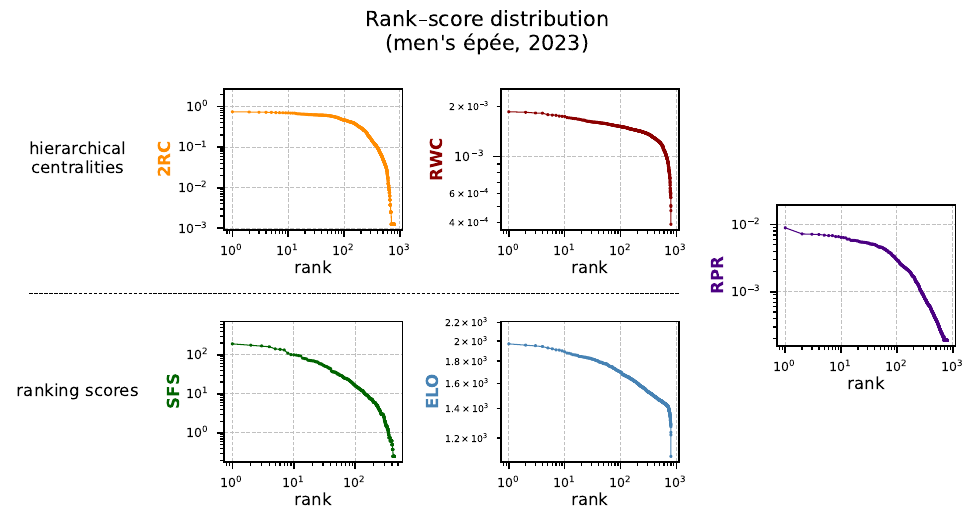}
    \caption{\textbf{The rank--score distribution of local measures for men's épée in 2023.} The rank--score distribution of nodes along the 2-reach centrality (orange), the random walk centrality (red), the official sport ranking (pink), the Elo score (cyan), and the reverse pagerank (purple) for the network of men's épée matches played in 2023. In the case of reversed pagerank two segments can be found following power-law scaling.}
\end{figure}

\begin{figure}[p]
    \centering
    \includegraphics[width=\textwidth]{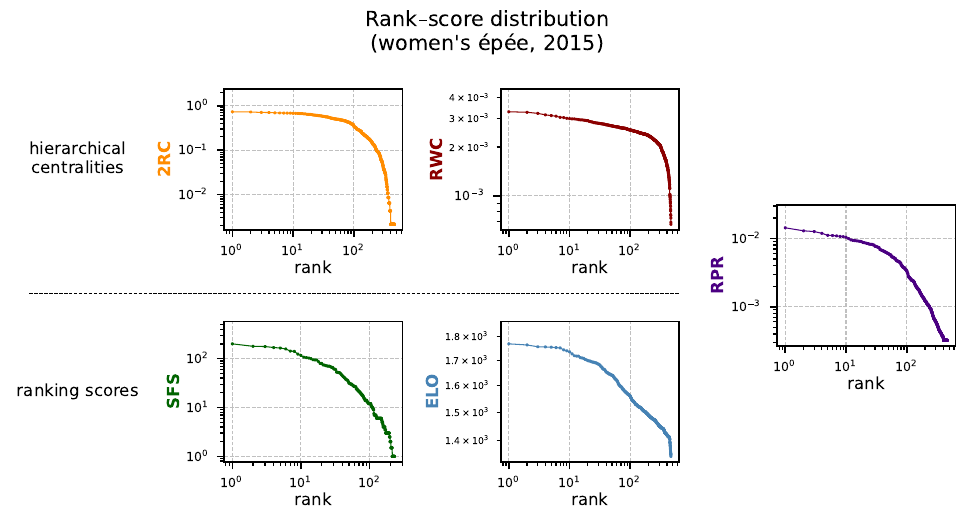}
    \caption{\textbf{The rank--score distribution of local measures for women's épée in 2015.} The rank--score distribution of nodes along the 2-reach centrality (orange), the random walk centrality (red), the official sport ranking (pink), the Elo score (cyan), and the reverse pagerank (purple) for the network of women's épée matches played in 2015. In the case of reversed pagerank two segments can be found following power-law scaling.}
\end{figure}

\begin{figure}[p]
    \centering
    \includegraphics[width=\textwidth]{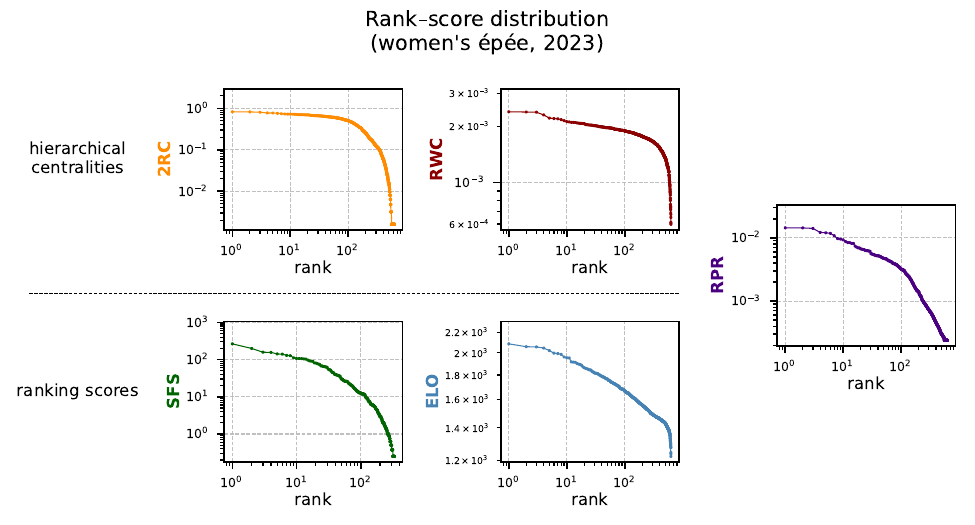}
    \caption{\textbf{The rank--score distribution of local measures for women's épée in 2023.} The rank--score distribution of nodes along the 2-reach centrality (orange), the random walk centrality (red), the official sport ranking (pink), the Elo score (cyan), and the reverse pagerank (purple) for the network of women's épée matches played in 2023. In the case of reversed pagerank two segments can be found following power-law scaling.}
\end{figure}

\begin{figure}[p]
    \centering
    \includegraphics[width=\textwidth]{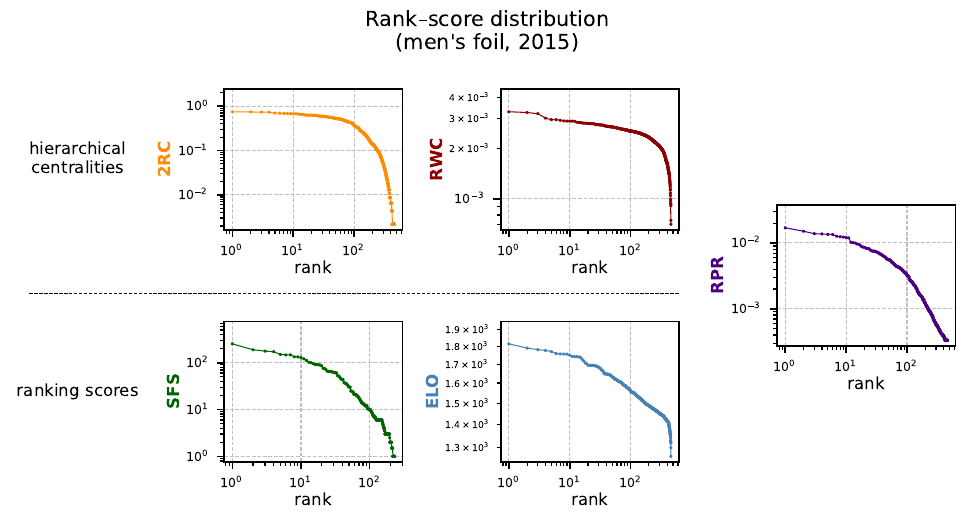}
    \caption{\textbf{The rank--score distribution of local measures for men's foil in 2015.} The rank--score distribution of nodes along the 2-reach centrality (orange), the random walk centrality (red), the official sport ranking (pink), the Elo score (cyan), and the reverse pagerank (purple) for the network of men's foil matches played in 2015. In the case of reversed pagerank two segments can be found following power-law scaling.}
\end{figure}

\begin{figure}[p]
    \centering
    \includegraphics[width=\textwidth]{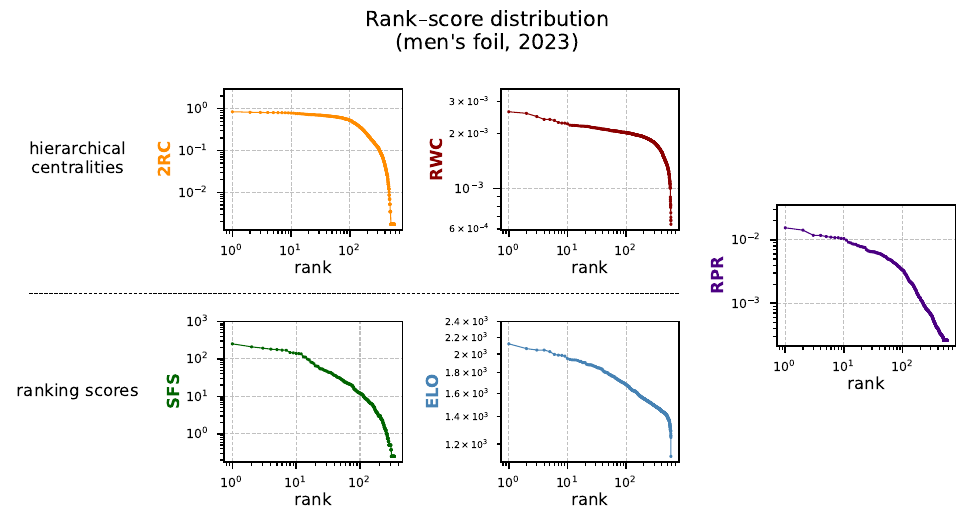}
    \caption{\textbf{The rank--score distribution of local measures for men's foil in 2023.} The rank--score distribution of nodes along the 2-reach centrality (orange), the random walk centrality (red), the official sport ranking (pink), the Elo score (cyan), and the reverse pagerank (purple) for the network of men's foil matches played in 2023. In the case of reversed pagerank two segments can be found following power-law scaling.}
\end{figure}

\begin{figure}[p]
    \centering
    \includegraphics[width=\textwidth]{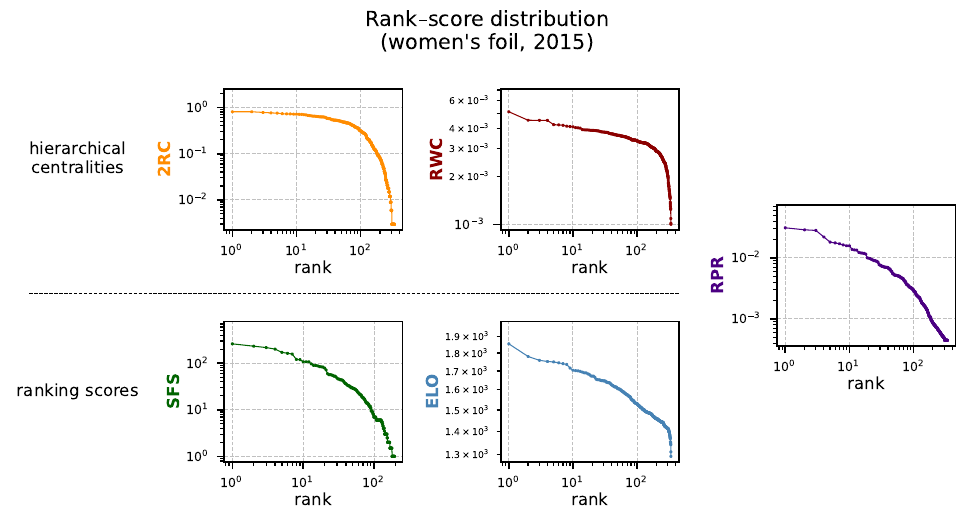}
    \caption{\textbf{The rank--score distribution of local measures for women's foil in 2015.} The rank--score distribution of nodes along the 2-reach centrality (orange), the random walk centrality (red), the official sport ranking (pink), the Elo score (cyan), and the reverse pagerank (purple) for the network of women's foil matches played in 2015. In the case of reversed pagerank two segments can be found following power-law scaling.}
\end{figure}

\begin{figure}[p]
    \centering
    \includegraphics[width=\textwidth]{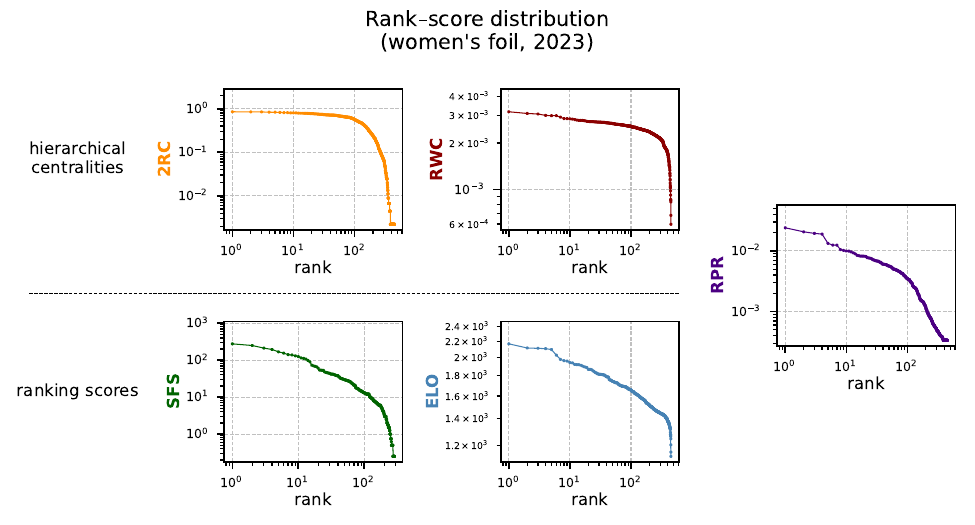}
    \caption{\textbf{The rank--score distribution of local measures for women's foil in 2023.} The rank--score distribution of nodes along the 2-reach centrality (orange), the random walk centrality (red), the official sport ranking (pink), the Elo score (cyan), and the reverse pagerank (purple) for the network of women's foil matches played in 2023. In the case of reversed pagerank two segments can be found following power-law scaling.}
    \label{fig:zipf-2}
\end{figure}

\end{document}